\documentstyle[eqsecnum,preprint,aps]{revtex}

\tighten

\begin{document}
\draft

\title{The energy dependence of the $\pi N$ amplitude and the three-nucleon
interaction}

\author{T-Y. Saito and I. R. Afnan}
\address{School of Physical Sciences,
        The Flinders University of South Australia,\\
        Bedford Park, SA 5042, Australia}
\date{\today}

\maketitle

\begin{abstract}

By calculating the contribution of the $\pi-\pi$ three-body force to the
three-nucleon binding energy in terms of the $\pi N$ amplitude  using
perturbation theory, we are able to determine the importance of the energy
dependence and the contribution of the different partial waves of the $\pi N$
amplitude to the three-nucleon force. A separable representation of the
non-pole
$\pi
N$ amplitude allows us to write the three-nucleon force in terms of the
amplitude for
$NN\rightarrow NN^*$, propagation of the $NNN^*$ system, and the amplitude for
$NN^*\rightarrow NN$, with $N^*$ being the $\pi N$ quasi-particle
amplitude in a given
state. The division of the $\pi N$ amplitude into a pole and non-pole gives a
procedure for the determination of the $\pi NN$ form factor within the
model. The total contribution of the three-body force to the binding energy
of the triton for the separable approximation to the Paris nucleon-nucleon
potential (PEST) is found to be very small mainly as a result of the energy
dependence of the $\pi N$ amplitude, the cancellation between the $S$- and
$P$-wave $\pi N$ amplitudes, and the soft $\pi NN$ form
factor.

\end{abstract}

\newpage

\section{INTRODUCTION}\label{sec.1}

The discrepancy between the results of the exact calculations for the binding
energy of the triton using a number of realistic nucleon-nucleon  potentials
and the experimental value of 8.48~MeV, has been an outstanding problem in
nuclear physics for a number of years~\cite{G93}. A commonly accepted
solution  has been the introduction of a three-nucleon force that will bridge
the gap between the calculated binding energy \cite{SI86,IS86,FG88,FPSd93},
based on a two-body interaction, and the experimental binding energy.
The origin of such a three-body force is partly the result of the  fact that
the nucleons are treated as point particles interacting via a two-body meson
exchange potential that is often assumed to be local.  The fact that the formal
division of the interaction into a two- and three-body part is not unique,
given the on-shell two-body data \cite{PG90}, suggests that the contribution
of the three-body force is partly determined by the definition of the
two-nucleon interaction. Thus under ideal conditions the division of the
interaction, in the three-nucleon system, between a two- and three-body force
will require a consistent formulation of these two potentials within a
meson-nucleon theory.

In the absence of such a formulation one may assume that a meson-nucleon
theory should give the correct binding energy for the three-nucleon system, in
which case the three-body force is by definition that force which when added
to the chosen two-nucleon force will give the three-nucleon binding
energy~\cite{CP83,SC86,Pa86}. A second approach is to assume that the
three-nucleon force is the result of meson exchanges that are possible only
when the number of nucleons is greater than two. In this second approach one
expects the dominant mechanism to be one in which one nucleon emits a meson
that scatters off a second nucleon and then gets absorbed on the third
nucleon, see Fig.~\ref{Fig.1}. In this case the three-nucleon force is
determined by the off-shell meson-nucleon amplitude that goes into the
calculation of the diagram in Fig.~\ref{Fig.1}. In the present investigation
we will consider the second approach involving a $\pi$ meson exchange. In
particular, we will examine the role of the energy dependence of this $\pi N$
amplitude on the contribution of this three-body force to the binding energy
of the triton. We will also examine the relative contribution of the different
$\pi N$ partial waves to this three-body force.

Over the past ten years three approaches have been developed  to
determine a three-nucleon interaction from $\pi N$ dynamics.
(i)~The Tucson-Melbourne (TM) \cite{Co79,CG81} three-nucleon potential is
based on the idea that the off-mass shell $\pi N$ amplitude should satisfy
current algebra constraints and the soft pion theorems~\cite{BG68}.  These
constraints allow a covariant parameterization of the off-mass shell amplitude.
To  be consistent with the meson exchange $NN$ interaction, the $\pi N$
amplitude is
expanded in powers of $\frac{q}{m_N}$, where $q$ is the pion momentum and $m_N$
the
nucleon mass. This gives a $\pi N$ amplitude that includes both $S$- and
$P$-wave
scattering, but where the energy dependence is reduced to $\nu=(s-u)/4m_N =
(q'+q)\cdot (p'+p)/4m_N=0$, as a result of the expansion in $q \over m_N$.
Here,
$s$
and $u$ are the usual Mandelstam variables while $q$ $(q')$ and $p$ $(p')$ are
the
initial (final) four momentum of the pion and nucleon respectively,  The
$\pi NN$ form factor is constructed  to satisfy the Goldberger-Treiman
relation\cite{GT58,CS90}.  Although the original TM potential included only the
$\pi-\pi$ three-body force, $\pi-\rho$ and $\rho-\rho$ contributions have
recently been
included~\cite{CP93}.  (ii)~A similar approach is to assume that the $\pi N$
dynamics
is determined by an effective chiral Lagrangian~\cite{CD83,RI84,RC86,CF86,We90}
which when used to calculate the $\pi N$ amplitude at the tree level, will
give an effective three-nucleon force given by the diagram in
Fig.~\ref{Fig.1}. The evaluation of the $\pi N$ amplitude at the tree level
gives rise to an energy independent $\pi N$ amplitude, and therefore a
three-body force. Both approaches (i) and (ii) give similar results, and
emphasize the chiral symmetry of the $\pi N$ amplitude.   (iii)~The $N-\Delta$
coupled channel approach~\cite{HSS83,HSY83,S86,PHS90,PRB91} takes as its
starting point the fact that the $\pi N$ amplitude is dominated at medium
energies by the $\Delta(1230)$ or the $P_{33}$ partial wave. This dominance of
the $\Delta$ suggests that we could extend our Hilbert space to include not
only the nucleon, but also the $\Delta$ as an excited state of the nucleon.
The approach of treating the $N$ and $\Delta$ on equal footing effectively
includes, in a consistent manner, that part of the original three-body force
corresponding to $\pi N$ scattering in the $P_{33}$ channel, or at least the
resonance part of it~\cite{SS92}. The advantage of this approach is that now
we can construct the two-body and the three-body forces with some
consistency, to the extent that the $NN-N\Delta$ transition potential used in
the two-body interaction can also be used to generate the three-body
potential. The inclusion of coupling between the $NN$, $N\Delta$ and
$\Delta\Delta$ channels allows a consistent treatment of the $BB$ and $BBB$
system, where $B=N,\Delta$\cite{PRB91}. In this approach, since the $\pi N$
amplitude is basically approximated by $\pi N\rightarrow\Delta\rightarrow\pi
N$, the energy dependence of the total $\pi N$ amplitude is completely
determined by the energy dependence of the $P_{33}$ channel which has the
$\Delta$ resonance. Although the $NN$ and transition potentials can be local
and energy independent, the effective three-body force in this model is
energy dependent and this energy dependence is determined by the $\Delta$
resonance, i.e. the $P_{33}$ amplitude. More recently there have been
extensions of the $N-\Delta$ coupled channel approach that have included the
$S$-wave component of the TM potential~\cite{SS92}. Also Pe\~na {\it et
al.}~\cite{PSSK93} have examined the importance of the coupling of the $\Delta$
to
the $\pi N$ channel. This latter calculation gives a $\pi N$ amplitude in
the $P_{33}$ channel that fits the phase shifts and has an
energy dependent mass and width for the $\Delta$. The inclusion of this
energy dependence in the $\Delta$ mass and width does not effect the final
result appreciably.

The questions which arise from the above approaches  to the $\pi N$ dynamics
that go into the derivation of the three-nucleon force and its contribution
to the binding energy of the three-nucleon system are:
(i)~What are the contributions of the different $\pi N$ partial waves to
the three-body force?
(ii)~If there is any cancellation between the contribution of the
different partial waves,
should the $P_{33}$ partial wave be treated via the $N-\Delta$
coupled channel while the rest of the $\pi N$ amplitude gives rise to a
three-body force as depicted in Fig.~\ref{Fig.1}?
(iii)~Would a cancellation between the different $\pi N$ channels be
sensitive to  the energy dependence of the amplitudes?
 The main aim of this investigation
is to examine these questions.

To motivate our interest in the importance of the energy dependence of
the $\pi N$ amplitude, let us examine the role of the $NN$ amplitude in
calculating the binding energy of the three-nucleon system within the
framework of the Faddeev equations. Here we observe that we require the fully
off-energy-shell $NN$ amplitude in a given partial wave $\alpha$,
$t_\alpha^{NN}(k,k';E_{NN})$, for all energies in the range
$-\infty<E_{NN}<-E_T$, where $E_T$ is the three-nucleon binding energy, see
Fig.~\ref{Fig.2}. The fact that we need this amplitude over the full
specified energy domain is a result of the fact that in the three-nucleon
system, the total energy is fixed at $E=-E_T$, and the spectator particle can
have any energy from zero to $\infty$. To consider the contribution of
Fig.~\ref{Fig.1} and in particular the energy dependence of the $\pi N$
amplitude, we consider the $NNN-\pi NNN$ equations\cite{AM83,CCS93}
which are an extension of the $NN-\pi NN$ equations to the $A=3$ sector.
Within the framework of the above $\pi NN$ dynamics, to calculate the
contribution of
the three-body force as defined in Fig.~1 to the three-nucleon binding, we must
determine to know the fully off-energy-shell $\pi N$ amplitudes \cite{Foot1},
$t_\alpha^{\pi N}(k,k';E_{\pi N})$, for all energies in the range
$-\infty<E_{\pi N}<(m_N-E_T)$, where in this case we have included the rest
mass of the nucleon and pion in the $\pi N$ energy $E_{\pi N}$. In other
words we have to calculate the fully off-energy-shell $\pi N$ amplitude for all
energies from $-\infty$ to $E_T$ below the nucleon pole ( see
Fig.~\ref{Fig.3}). Thus to calculate a three-body force which is defined in
terms of the $\pi N$ amplitude, we need to know this amplitude off-shell for
energies far below ($(m_\pi + E_T)\approx 145$~MeV), the $\pi N$ threshold.
Here the need for the $\pi N$ amplitude over this energy domain is a result
of the fact that the total energy is still $-E_T$, but now we have two
spectator nucleons with kinetic energy between zero and $\infty$. At these
energies it is not clear that the $\Delta(1230)$ is dominant or that the
$P$-wave amplitudes are more important than the $S$-wave amplitudes.
Clearly, the threshold behavior of the amplitude is essential, but we need to
know the
$\pi N$ partial wave amplitudes over a wide range of energies above
threshold, if we are to extrapolate these amplitudes to the energies required
in calculating the three-body force. This suggests that we have to fit the
energy dependence of the experimental $\pi N$ phase shifts in order to improve
the
accuracy of the extrapolation in the energy.

In Sec.~\ref{sec.a} we will  derive  the three-body force using
the coupled channel method and we will discuss approximation which we
take to avoid the dressing problem\cite{SS85}.
In Sec.~\ref{sec.2} we will discuss a parameterization for the $\pi N$
amplitude, and
in particular will discuss the division of this amplitude into a pole and
non-pole component. It is only the non-pole part of the
$\pi N$ amplitude that goes into the $\pi N$ scattering that generates the
three-body force. This division is essential if we are to avoid double
counting.  We will choose the $\pi NN$ form factor as the residue of the pole
term
in the $\pi N$ amplitude  to keep the consistency with the $\pi N$ formulation.
We will then proceed in Sec.~\ref{sec.3} to a discussion of the $\pi-\pi$
three-body
force given in Fig.~\ref{Fig.1} in terms of our parameterization of the $\pi N$
amplitude. As a result of using a separable representation for the $\pi N$
amplitude,
we find that the three-body force is reduced to the product of the amplitude
for
$NN\rightarrow NN^*_\alpha$, followed by the propagation of the $NNN^*_\alpha$
system, and the final transition amplitude for
$NN^*_\alpha\rightarrow NN$,  where $\alpha$ runs over all $\pi N$ partial
waves.
This $\pi-\pi$ three-body force is employed in Sec.~\ref{sec.4}  to calculate,
in the
Born approximation, the contribution of this force to the binding energy of
the three-nucleon system. The three-nucleon wave function will be calculated by
using
Paris (PEST)  potential\cite{HP84}. Since the Paris potential does not have any
energy  dependence, our approach for the three-nucleon interaction is
inconsistent
with the two-nucleon interaction. To that extent, the overall magnitude of
the resultant three-body force contribution may not have great significance.
However, we would like to examine the relative contribution of the
different $\pi N$ partial waves, and see how this contribution is sensitive to
the
energy dependence of the $\pi N$ amplitude in the subthreshold region.
In particular, we would like to demonstrate how the energy dependence of both
the
non-pole amplitude and the $\pi NN$ form factor play a role in the importance
of
this
three-nucleon force. Here we will find that the energy dependence of the
non-pole
amplitude has a significant effect on the contribution of the three-body force.
Furthermore, we will demonstrate that there is a cancellation between the
contribution of the different $\pi N$ partial waves. This cancellation turns
out
to
be very sensitive to the approximations used. Finally, in Sec.~\ref{sec.5} we
will
conclude our discussion by considering some open questions that can influence
our
final results.

\section{COUPLED CHANNEL FORMULATION OF THREE-BODY FORCE}\label{sec.a}

In this section, we will establish the approximation involved
in writing Fig.~\ref{Fig.1} as the lowest order contribution
to the binding energy of the three-nucleon system from a
$\pi$-$\pi$ three-body force. Our starting point is the Hamiltonian
of Mizutani and Koltun\cite{MK77}
\begin{equation}
{\cal H} = {\cal K} + {\cal V}_{NN} + {\cal V}_{\pi N}
+ {\cal A} + {\cal A}^\dagger\ ,                                 \label{eq:1.1}
\end{equation}
where ${\cal K}$ is the kinetic energy of the nucleons and pions, while  ${\cal
V}_{NN}$ and ${\cal V}_{\pi N}$ are the $NN$ interaction in the absence  of one
pion
exchange, and the $\pi N$ interaction in the absence of the s-channel nucleon
pole,
respectively. In Eq.~(\ref{eq:1.1}),  ${\cal A}$ is the pion absorption vertex
while
${\cal A}^\dagger$ is the corresponding  production vertex.  The Schr\"odinger
equation for this Hamiltonian is
\begin{equation}
{\cal H}  |\Psi \rangle = E \, |\Psi \rangle\ .                 \label{eq:1.2}
\end{equation}
Clearly, the operators ${\cal A}$ and ${\cal A}^\dagger$ change
the number of pions. To that extent, the present Hamiltonian has some
of the features of a field theory in that the number of pions is not fixed
while
the
number of nucleons is fixed.  As a first approximation, we restrict the Hilbert
space
to $(nN)$ and $\pi (nN)$ systems only, where $n$ is the number of nucleons,
which is
conserved. We define the Feshbach~\cite{Fe62} projection operators onto $(nN)$
and
$\pi (nN)$ spaces as  $P$ and $Q$, respectively. By using these projection
operators
and by assuming that this truncated space is complete, {\it i.e.} $P + Q = 1$,
the wave
functions of the $(nN)$ and $\pi (nN)$ components (~$P\Psi$ and $Q\Psi$,
respectively)
are solutions of the equations
\begin{equation}
\Bigl(E - H_{PP} - H_{PQ} {1\over E-H_{QQ}}
H_{QP} \Bigr) | P \Psi \rangle = 0\ ,                           \label{eq:1.3}
\\
\end{equation}
and
\begin{equation}
\Bigl(E - H_{QQ} - H_{QP} {1\over E-H_{PP}}
H_{PQ} \Bigr) | Q \Psi \rangle = 0\ .                          \label{eq:1.4}
\end{equation}
In writing Eqs.~(\ref{eq:1.3}) and (\ref{eq:1.4}) we have made use of
the definitions,
\begin{eqnarray}
H_{PP} \equiv P{\cal H}P\ ,\quad &\quad&\quad H_{PQ} \equiv P{\cal H}Q\ ,
\nonumber \\
H_{QP} \equiv Q{\cal H}P\ ,\quad &\mbox{and}&\quad H_{QQ} \equiv Q{\cal H}Q
\ .                                                            \label{eq:1.5}
\end{eqnarray}
The Green's function for the $\pi(nN)$ system, $(E - H_{QQ})^{-1}$, which
including
the $\pi N$ and $NN$ interactions,  can be written as
\begin{equation}
{1\over E - H_{QQ}} = G^{(0)}_{\pi (nN)}(E) +
G^{(0)}_{\pi (nN)}(E) \,T(E) \, G^{(0)}_{\pi (nN)}(E)\ ,        \label{eq:1.6}
\end{equation}
where $G^{(0)}_{\pi (nN)}(E)$ is the free $\pi (nN)$ propagator.  The
corresponding
T-matrix, $T(E)$, can be written as
\begin{equation}
T(E) = \sum_a \, t_a(E) + \sum_{ab} \,
t_a(E) \, G^{(0)}_{\pi (nN)}(E) \, U_{ab}(E) \, G^{(0)}_{\pi (nN)}(E)
\, t_b(E)\ ,                                                     \label{eq:1.7}
\end{equation}
where $U_{ab}(E)$ is the AGS amplitude\cite{AGS67} for the $(n+1)$
particle system and satisfies a set of coupled equations. Here, $t_a(E)$
is the $\pi N$ or $NN$ amplitudes resulting from the potential ${\cal V}_{NN}$
or
${\cal V}_{\pi N}$, respectively.

We will first consider the simplest case of $n=1$ \cite{EA89}. Here, on the one
hand,
Eq.(\ref{eq:1.3}) can be regarded as the equation for the dressed nucleon with
the
self-energy, $\Sigma(E)$, defined to be
\begin{equation}
\Sigma(E) = H_{PQ} \, {1\over E - H_{QQ}} \, H_{QP}\ .
\label{eq:1.8}
\end{equation}
Therefore the dressed nucleon propagator, $G_N(E)$, can be written as
\begin{equation}
G_N(E) = \left(E - H_{PP} - H_{PQ} \, {1 \over E - H_{QQ} } \,
H_{QP} \right) ^{-1}\ ,
\label{eq:1.9}
\end{equation}
with $H_{PP}$, the bare mass of the nucleon. On the other hand,
Eq.~(\ref{eq:1.4})
describes $\pi N$ scattering. Now since $H_{QQ}= K_{QQ} + {\cal V}_{\pi N}$,
with
$K_{QQ}$ the kinetic energy of the $\pi N$ system, the $\pi N$ interaction in
Eq.~(\ref{eq:1.4}) is the sum of two contributions. The first is the $\pi N$
interaction in the original Hamiltonian, ${\cal V}_{\pi N}$, while the second
term,
$H_{QP}\left(E-H_{PP}\right)^{-1}H_{PQ}$, results from the coupling of the
$\pi N$ channel to the $N$ channel. The $\pi N$ T-matrix, $T_{\pi N}(E)$, in
this
case is the  sum of two terms, and is given by
\begin{equation}
T_{\pi N}(E) = F_{QP}(E) \, G_N(E) \, F_{PQ}(E) + t_B(E)\ ,     \label{eq:1.10}
\end{equation}
where
\begin{equation}
F_{PQ}(E)=H_{PQ}\left[1+\frac{1}{E - K_{QQ}}\,t_B(E)\right]\ ,  \label{eq:1.11}
\end{equation}
and $t_B(E)$ is the solution to the two-body equation
\begin{equation}
t_B(E) = V_{QQ} + V_{QQ} \, \frac{1}{ E - K_{QQ}} \, t_B(E)\ ,  \label{eq:1.12}
\end{equation}
with $V_{QQ}={\cal V}_{\pi N}$. Here we note that the first term on the right
hand
side of Eq.~(\ref{eq:1.10}) corresponds to the process $\pi N\rightarrow
N\rightarrow
\pi N$, and as a result the pion absorption (emission) vertex is given by
$F_{PQ}(E)$  ($F_{QP}(E)$). To establish that this term has the nucleon pole
contribution to the $\pi N$ amplitude, we note that the Green's function,
$G_N(E)$,
given in Eq.~(\ref{eq:1.9}), has a pole at the nucleon mass, and therefore can
be
written as
\begin{equation}
G_N(E)=\frac{1}{E-H_{PP}-\Sigma(E)}=\frac{Z_2(E)}{E-m_N}\ ,     \label{eq:1.13}
\end{equation}
where $Z_2(E)$ is the wave function renormalization. Making use of this result
in
Eq.~(\ref{eq:1.10}) allows us to write the $\pi N$ T-matrix as
\begin{equation}
T_{\pi N}(E) = f^{R\dag}_{\pi NN}(E)\,\frac{1}{E-m_N}\,f^{R}_{\pi NN}(E)
                 + t_B(E)\ ,                                    \label{eq:1.14}
\end{equation}
where the renormalized $\pi NN$ form factor, which is energy dependent, is
given
by
\begin{equation}
f^R_{\pi NN}(E) = Z^{1\over2}_2(E) \, F_{PQ}(E)\ .
\label{eq:1.15}
\end{equation}
In this formulation the normalized physical nucleon wave function, which is a
solution to Eq.~(\ref{eq:1.3}) is given by
\begin{equation}
\Psi_N = Z^{\frac{1}{2}}_2(m_N)\,P\Psi\ .
\label{eq:1.16}
\end{equation}

We consider next the case of $n>1$. Since we have restricted the Hilbert space
to
include $n$ nucleons, and up to one pion only, this truncation effects the
dressing of
the nucleons for the case $n>1$. From Fig.~\ref{Fig.4}, we observe that the
nucleons
can only be dressed separately after absorbing an initial pion and before
emitting the
final pion, because of the limitation imposed on the  Hilbert space. We can
neither
include the nucleon dressing before the pion absorption nor after the final
pion
emission. This incomplete  dressing makes the renormalized $\pi NN$ form
factor,
$f^R_{\pi NN}(E)$, smaller\cite{SS85}. To overcome this problem we need to
guarantee
that all the nucleons are fully dressed at the same time, and this dressing is
on both
sides of the $\pi NN$-vertex\cite{KB94}, see Fig.~\ref{Fig.5}.  We will avoid
this
problem in the present investigation in the following manner.  In the $NN-\pi
NN$
equations the nucleon dressing was introduced to satisfy the unitarity of the
$NN$
amplitude above the pion production threshold. In the present investigation
however,
we are  considering a bound state problem which is below the  threshold for
pion
production. We therefore expect that the nucleon dressing will not be
essential,
and
does not have to be included explicitly to satisfy unitarity. We therefore
assume that
each nucleon line has been renormalized to give the physical nucleon mass and
correct
$\pi NN$ coupling constant, {\it i.e.}; (i)~The nucleon propagator is  given by
$(E-m_N)^{-1}$. (ii)~The $\pi NN$ vertex is given by the renormalized vertex
function,
$f^R_{\pi NN}(E)$.

Using the above assumptions, the interaction term resulting from the coupling
of
the
$NNN$ and $\pi NNN$ Hilbert space in Eq.~(\ref{eq:1.3}) can be rewritten  as
\begin{equation}
H_{PQ} \, {1\over E - H_{QQ}} \, H_{QP} = {\cal V}_{OPE}
+ {\cal V}_{disp} + {\cal V}_{3B} + \cdots\ ,
\label{eq:1.17}
\end{equation}
where the series is generated by iterating the equation for $U_{ab}$ in
Eq.~(\ref{eq:1.7}), and making use of the resultant $\pi NNN$ T-matrix in
Eq.~(\ref{eq:1.6}). In particular, if we go to third order in the
$\pi N$  amplitude $t_a$, the $\pi NN$ form factors in ${\cal V}_{3B}$ get
dressed
as detailed in Eqs.~(\ref{eq:1.14}) and (\ref{eq:1.15}).  Each of the terms in
Eq.~(\ref{eq:1.17}) is shown schematically in Fig.~\ref{Fig.6}.  We note that
${\cal
V}_{OPE}$ and ${\cal V}_{disp}$ are two-body  operators. Since we will be
calculating
perturbatively the contribution of the three-body force to the binding energy
of
$^3$H,  the three nucleon wave function used results from the solution of the
Faddeev
equations for a given nucleon-nucleon interaction. Assuming that the
nucleon-nucleon
interaction includes ${\cal V}_{OPE}$ and ${\cal V}_{disp}$, we omit these
terms
from
Eq.~(\ref{eq:1.17}).  This leaves us with the three-nucleon force ${\cal
V}_{3B}$,
illustrated in  Figs.~\ref{Fig.6}(c) and \ref{Fig.1}, which gives the lowest
order
correction  to the binding energy of the three-nucleon system. The detail of
the
numerical framework of the $\pi N$  scattering will be presented in next
section.

\section{THE $\pi N$ AMPLITUDE}\label{sec.2}

To include the full energy dependence of the $\pi N$ amplitude into the
determination of the $\pi-\pi$ three-nucleon interaction and its contribution
to the binding energy of the triton, we have to:
(i)~Remove the nucleon pole contribution from the $\pi N$ amplitude to avoid
double counting in the three nucleon force calculation.
(ii)~Define a $\pi NN$ form factor for the emission and absorption vertices in
Fig.~\ref{Fig.1} that is consistent with the $\pi N$ amplitude used and
the  scattering data.
(iii)~Treat the nucleon in the $\pi N$ system using non-relativistic
kinematics to maintain consistency between the $NNN$ and $\pi N$ systems since
the nucleons in the three-nucleon system are treated non-relativistically.
{}From the previous section we observe
the first two conditions can be satisfied if we choose a formulation of the
$\pi N$ scattering problem that is motivated by a Hamiltonian that includes a
$\pi NN$ vertex and a $\pi N$ interaction, e.g. the Cloudy Bag
Model~\cite{Th84}. Thus a choice for the $\pi N$ potential, motivated by the
lowest order contribution to the amplitude and based on a Lagrangian of the
form suggested by the Cloudy Bag Model with volume coupling~\cite{Th81},
consists of an
$s$-channel nucleon pole diagram, Fig.~\ref{Fig.7}, and the cross diagram
and contact term, Fig.~\ref{Fig.8}.

For the present investigation, to simplify the parameterization of the  $\pi
N$ amplitude, we replace the cross diagram and contact term in
Fig.~\ref{Fig.8} by a one term separable potential in each partial wave. This
allows us to write the $\pi N$ potential in a given partial wave $\alpha$ as
\begin{equation}
v_\alpha(k,k';E) = f_0(k)\,\frac{1}{E-m_0}\,f_0(k') +
g_\alpha(k)\,\lambda_\alpha\,g_\alpha(k')\ ,                  \label{eq:2.1}
\end{equation}
where the first term corresponds to the nucleon pole diagram with a bare
nucleon mass of $m_0$, and a bare form factor $f_0(k)$. Since we
are using non-relativistic kinematics for the nucleon, this $s$-channel pole
diagram contributes to the $P_{11}$ partial wave only. As a result, the
potential in all partial waves other than the $P_{11}$ channel is given by
the second term on the right hand side of Eq.~(\ref{eq:2.1}). This separable
$\pi N$ potential  gives an amplitude of the form
\begin{equation}
t_\alpha(k,k';E) = g_\alpha(k)\,\tau_\alpha(E)\,g_\alpha(k')
\quad\mbox{for}\quad \alpha\neq P_{11}                        \label{eq:2.2}
\end{equation}
where
\begin{equation}
\tau_\alpha(E) = \left[\lambda_\alpha^{-1} -
\langle g_\alpha|G_{\pi N}(E)|g_\alpha\rangle\right]^{-1}\ ,  \label{eq:2.3}
\end{equation}
with
\begin{equation}
\langle g_\alpha|G_{\pi N}(E)|g_\alpha\rangle = \int\limits^\infty_0 \,dk\,k^2
\frac{[g_\alpha(k)]^2} {E- \omega_k - \frac{\textstyle{k^2}}
{\textstyle{2m_N}} - m_N}\ .                                  \label{eq:2.4}
\end{equation}
This amplitude is a solution of the Lippmann-Schwinger equation with the pion
energy treated relativistic since $\omega_k = \sqrt{k^2 + m_\pi^2}$. The
strength of the potential $\lambda_\alpha$, and the form factor $g_\alpha(k)$
are
adjusted to fit the experimental $\pi N$ phase shifts in all $S$- and
$P$-waves other than the $P_{11}$ channel. For the present investigation we
will use the parameterization used by Thomas~\cite{Th76} for $\pi d$
scattering.
The form factor used for $S_{11}$ and $S_{31}$ $\pi N$ partial waves is:
\begin{equation}
g_\alpha(k) = {S_1 \over k^2 + \alpha_1^2} + {S_2 \over k^2 + \alpha_2^2},
\label{eq:2.a1}
\end{equation}
and
\begin{equation}
g_\alpha(k) = k \Biggl[{S_1\over k^2 + \alpha_1^2} +
{S_2 \, k^2 \over (k^2 + \alpha_2^2)^2 } \Biggr]
\label{eq:2.a2}
\end{equation}
for $P_{13}$, $P_{31}$ and $P_{33}$.
This parameterization has also been used extensively in the
$NN-\pi NN$ calculation for $\pi d$ scattering and $\pi d\rightarrow pp$
reactions~\cite{BA81,AM85}.

The $P_{11}$ channel plays the important role in this analysis of the
three-nucleon interaction as it has the nucleon pole contribution that needs
to be removed to avoid double counting. It also has the information about the
$\pi NN$ form factor which is defined as the residue of the off-shell $\pi N$
amplitude at the nucleon pole. In this way we can subtract the nucleon pole
contribution to the $\pi N$ amplitude, and extract a $\pi NN$ form factor
while maintaining a fit to the $\pi N$ scattering data in this channel. Since
the potential in this channel is the sum of two contributions, see
Eq.~(\ref{eq:2.1}), we can write the corresponding amplitude as a solution of
the Lippmann-Schwinger equation using a two-potential theory to be~\cite{AS81}
\begin{equation}
t_\alpha(k,k';E) = f(k;E)\,\frac{1}{E-m_0-\Sigma(E)}\,f(k';E) +
g_\alpha(k)\,\tau_\alpha(E)\,g_\alpha(k')\ ,                  \label{eq:2.5}
\end{equation}
where the dressed $\pi NN$ form factor, $f(k;E)$, is given by
\begin{equation}
f(k;E) = f_0(k) + g_\alpha(k)\,\tau_\alpha(E)\,
\langle g_\alpha|G_{\pi N}(E)|f_0\rangle\ ,                    \label{eq:2.6}
\end{equation}
and $\alpha$ in Eq.~(\ref{eq:2.5}) and (\ref{eq:2.6}) refers to the
$P_{11}$ channel. In Eq.~(\ref{eq:2.5}), the mass renormalization factor
$\Sigma(E)$ is given by
\begin{eqnarray}
\Sigma(E) &=& \langle f_0|G_{\pi N}(E)|f(E)\rangle\nonumber \\
          &=& \langle f_0|G_{\pi N}(E)|f_0\rangle +
              \langle f_0|G_{\pi N}(E)|g\rangle\,\tau(E)\,
              \langle g|G_{\pi N}(E)|f_0\rangle \ ,            \label{eq:2.7}
\end{eqnarray}
where we have dropped the channel label $\alpha$ with the understanding that
we are considering the $P_{11}$ partial wave only.

In writing the amplitude in the $P_{11}$ channel ( Eq.~(\ref{eq:2.5}) ) as
the sum of a pole  and a background term, we are able to define that part of
the $\pi N$ amplitude which will be included in the evaluation of the
three-body force. At the same time, we can determine the $\pi NN$ form factor
that is required for the pion emission and absorption vertices in
Fig.~\ref{Fig.1}. To establish that this $\pi NN$ form factor gives the correct
$\pi NN$ coupling constant as the residue of the $\pi N$ amplitude at the
nucleon pole, we expand $\Sigma(E)$ about the physical nucleon mass
as~\cite{MA85}
\begin{equation}
\Sigma(E) = \Sigma(m_N) + (E-m_N)\,\Sigma_1(m_N) +
           (E-m_N)^2\,\Sigma_2(E)\ .                        \label{eq:2.8}
\end{equation}
If we now fix the bare mass $m_0$ such that
\begin{equation}
m_0 + \Sigma(m_N) = m_N      \ ,                            \label{eq:2.9}
\end{equation}
we can write the pole amplitude as
\begin{equation}
f(k;E)\,\frac{1}{E-m_0 -\Sigma(E)}\,f(k';E) =
f^R(k;E)\,\frac{1}{E - m_N}\,f^R(k';E)                      \label{eq:2.10}
\end{equation}
where the renormalized $\pi NN$ form factor $f^R(k;E)$ is defined as
\begin{equation}
f^R(k;E) = \frac{Z_2^{1/2}}
          {[1 - (E-m_N)\Sigma^R_2(E)]^{1/2}}\,f(k;E)\ ,     \label{eq:2.11}
\end{equation}
and the wave function renormalization constant $Z_2$ is given by
\begin{equation}
Z_2 = \left[1 - \Sigma_1(m_N)\right]^{-1}
   \equiv 1 + \Sigma^R_1(m_N)\ ,                            \label{eq:2.12}
\end{equation}
with
\begin{equation}
\Sigma_i^R(E) \equiv Z_2\, \Sigma_i(E)\quad\mbox{for}\quad i=1,2\ .
                                                              \label{eq:2.13}
\end{equation}
In this way we have defined the renormalized $\pi NN$ form factor, $f^R(k;E)$,
which will be used for the emission and absorption vertices in
Fig.~\ref{Fig.1}. More important is the fact that this form factor is
constrained by the $\pi N$ phase shifts in the $P_{11}$ channel and the
requirement that we have the correct $\pi NN$ coupling constant. We note at
this point that the renormalized $\pi NN$ form factor $f^R(k;E)$ is energy
dependent and that this energy dependence is determined by unitarity through
$\tau_\alpha(E)$, and has to be included if we are to fit the phase shifts in
this channel.

To determine the $\pi NN$ coupling constant resulting from the above
formulation of $\pi N$ scattering, we need to compare our results for the pole
amplitude with the corresponding Feynman diagram for the Lagrangian with the
pseudoscalar coupling, i.e.
\begin{equation}
-ig_0(k)\bar{u}\,(\bbox{\tau}\cdot\bbox{\phi})\,(i\gamma_5)\,u\ .
                                                             \label{eq:2.14}
\end{equation}
where the coupling constant $g_0$ is made a function of the momentum. In
Eq.~(\ref{eq:2.14}), $\bbox{\tau}$ is the Pauli isospin matrix, $\bbox{\phi}$
is the pion field, and $u$ is the usual Dirac spinor. This interaction
Lagrangian allows us to determine the invariant amplitude corresponding to the
$s$-channel nucleon pole diagram~\cite{BD64}, and the corresponding
$S$-matrix. Making use of the relation between the $S$-matrix and the
$T$-matrix~\cite{AT76}, which is a solution of the Lippmann-Schwinger
equation, we can calculate the $\pi NN$ coupling constant as a result of the
relation between the $T$-matrix, $t(k,k';E)$, and the invariant amplitude
${\cal T}(k,k';E)$ as
\begin{equation}
t(k,k';E) = C(k;E)\,{\cal T}(k,k';E)\,C(k';E)              \label{eq:2.15}
\end{equation}
where
\begin{equation}
C(k;E) = \sqrt{\frac{m_N(\omega_k +
\varepsilon_k)}{(2\pi)^3\,\omega_k\varepsilon_k(E + \omega_k + \varepsilon_k)}}
\ ,                                                         \label{eq:2.16}
\end{equation}
with $\varepsilon_k = \sqrt{k^2 + m_N^2}$. The coupling constant is now
defined as the residue of the invariant $\pi N$ amplitude at the nucleon pole
with all the legs of the $\pi NN$ vertex on-mass-shell. This corresponds to
taking $E=m_N$ and $k=k_0$, where
\begin{equation}
k_0^2 = - m_\pi^2\left(1 - \frac{m_\pi^2}{4m_N^2}\right)\ .   \label{eq:2.17}
\end{equation}
This definition allows us to write the $\pi NN$ coupling constant $f_{\pi
NN}$ in terms of the renormalized $\pi NN$ form factor $f^R(k;E)$ as
\begin{eqnarray}
f^2_{\pi NN}(k) &=& \frac{m_\pi^2}{4m_N^2}\ g_0^2(k)
\nonumber \\
  &=& \frac{m_\pi^2}{4 m_N^2}\,\frac{m_N(\varepsilon_k+ m_N)}
{6\pi [C(k,m_N)]^2}
\ \left[\frac{f^R(k;m_N)}{k}\right]^2         \ ,           \label{eq:2.18}
\end{eqnarray}
where
\begin{equation}
\frac{1}{4\pi}\,f_{\pi NN}^2(k_0) = 0.079\   .              \label{eq:2.19}
\end{equation}
For the present investigation we make use of the $P_{11}$ $\pi N$
parameterization of McLeod and Afnan~\cite{MA85}, where the bare $\pi NN$ form
factor $f_0(k)$ is taken to be
\begin{equation}
f_0(k) = \frac{c_0}{\sqrt{\omega_k}}\
\frac{k}{(k^2 + \alpha^2)^{n_0}} \ ,                       \label{eq:2.20}
\end{equation}
while for the background separable potential form factor, we take
\begin{equation}
g(k) = \frac{k}{\sqrt{\omega_k}}\ \left[\frac{c_1}{k^2 + \beta_1^2} +
\frac{c_2 (k^2)^{n_1}}{(k^2 + \beta_2^2)^3}\right]\ .         \label{eq:2.21}
\end{equation}
This choice of form factor is basically the same as that used by Thomas
( Eqs.(\ref{eq:2.a1}) and (\ref{eq:2.a2}) ). The factor of $\sqrt{\omega_k}$
was introduced to get the covariant phase space to determine the
coupling constant at the nucleon pole with all legs on-mass-shell.
In Table~\ref{Table.1} we present two parameterizations of the $P_{11}$
amplitude~\cite{MA85} corresponding to a monopole ( $n_0=1$ ) or a dipole (
$n_0=2$ ) bare $\pi NN$ form factor. The parameters  were adjusted to give the
phase shifts below the pion production threshold, the position of the nucleon
pole, and the $\pi NN$ coupling constant of 0.079. In Table~\ref{Table.2} we
present the scattering volume $a_{11}$ in this channel, the wave function
renormalization $Z_2$ and the value of the coupling constant at $k=0$. We note
that the values in Tables~\ref{Table.1} and \ref{Table.2} are the corrected
values for the parameters of these potentials and their predictions.
We note that the
renormalized form factor $f^R(k;E)$ is substantially different from the bare
form factor $f_0(k)$ due to the wave function renormalization $Z_2$
and the contribution of the non-pole amplitude to the form factor dressing
(see Eq.(\ref{eq:2.6}) and (\ref{eq:2.11})). Finally, the value of
the coupling constant at $k=0$, when compared with the value at $k=k_0$, can
be used as a measure of the deviation from the Goldberger-Treiman relation.

The present definition of the $\pi NN$ form  factor is different from that
commonly used in $NN$ potentials, and three-nucleon force calculations.
However it is consistent with the formulation of $\pi N$ scattering where
the $\pi N$ amplitude is a solution to a two-body equation\cite{PJ91}.
Traditionally the $\pi NN$ form factor, introduced as a cutoff in the
$NN$ amplitude, is a function of the pion momentum only. This is a result of
taking both nucleons in the vertex on-mass-shell. However in a
non-relativistic or time ordered theory, intermediate particles are off the
energy shell. As a result, the $\pi NN$ form factor becomes
a function of the energy and the relative momentum. The energy dependence of
the
$\pi NN$ form factor is the result of the dressing. This dressing is
necessitated by
the requirement that the full $\pi N$ amplitude, even in the $P_{11}$ channel,
should
be a solution of the Lippmann-Schwinger equation. In this way we can maintain
consistency with both of the treatments for the $\pi N$ and $NN$ amplitudes  as
solutions of two-body equations when used in the three-nucleon system.

\section{THE THREE-BODY FORCE}\label{sec.3}

Having defined our $\pi N$ amplitude, and in particular how the  nucleon pole
is subtracted from the $\pi N$ amplitude to give us a $\pi NN$ form factor
that is both momentum and energy dependent, we turn our attention to our
definition of the three-body force as given in Fig.~\ref{Fig.1}. At this stage
it is important to note that this definition of the three-body force does not
include
all possible pion exchange diagrams that are not included in the
nucleon-nucleon
interaction. However, we expect the diagram in  Fig.~\ref{Fig.1} to give the
dominant
contribution to the three-body force.

To evaluate the diagram in Fig.~\ref{Fig.1} we introduce Jacobi  variables in
the $\pi NNN$ center of mass. These are defined in Fig.~\ref{Fig.9}. This
choice for the momenta will allow us to take matrix elements of the three-body
force between three-nucleon wave functions resulting from the solution of the
Faddeev equation in momentum space for a given two-nucleon interaction. At the
same time we will be able to include the $\pi N$ amplitudes defined in
Sec.~\ref{sec.2} with their full energy dependence with no approximations. The
momenta in the initial state in Fig.~\ref{Fig.9} are
\begin{eqnarray}
{\bf q}_3 &=& - {\bf k}_3                                    \label{eq:3.1} \\
{\bf p}_3 &=&  \frac{m_N ({\bf k}_\pi + {\bf k}'_1) -
(m_N+m_\pi){\bf k}_2 }{(2m_N + m_\pi)}                       \label{eq:3.2} \\
{\bf Q}_3 &=& \frac{m_N {\bf k}_\pi - m_\pi {\bf k}'_1 }
{ (m_N + m_\pi)}\ ,                                          \label{eq:3.3}
\end{eqnarray}
where ${\bf k}'_1$, ${\bf k}_2$, ${\bf k}_3$ and  ${\bf k}_\pi$ are the
momenta of the three nucleon and pion after the pion emission vertex. Here,
${\bf Q}_3$ is the $\pi N$ relative momentum for the pion production vertex,
while ${\bf Q}'_3$ is the relative $\pi N$ momentum in the $\pi N$ amplitude.
In a similar manner we can define all the Jacobi momenta before and after the
$\pi N$ scattering and before the pion absorption. In this way all momenta
 are defined in terms of the initial and final momenta of the three
nucleons. At this stage we should point out that in a non-relativistic theory,
the $\pi NN$ vertex is not Galilean invariant since mass is not conserved. As
a result, the relative momenta ${\bf p}_3$ and ${\bf q}_3$ are not the same
before and after the pion emission. For practical calculations we will assume
that not to be the case. In other words the relative momenta ${\bf p}_3$ and
${\bf q}_3$ are those used in the three-nucleon wave function resulting from
the solution of the Faddeev equations.

The pion absorption and emission vertices in Fig.~\ref{Fig.9}  have the form
factor $f^R(Q_i;E_i), i=1,3$ respectively, where the energy $E_i$ is the
energy available to the $\pi N$ system, and is given by
\begin{equation}
E_i = E + m_N - \frac{q_i^2}{2\mu_2} - \frac{p_i^2}{2\mu_1}\ , \label{eq:3.4}
\end{equation}
where $E = - E_T$ is the total energy of the system not including rest
masses, and the reduced masses $\mu_1$ and $\mu_2$ are defined by the relations
\begin{equation}
\frac{1}{\mu_1} = \frac{1}{m_N} + \frac{1}{m_N + m_\pi}\ , \quad\mbox{and}\quad
\frac{1}{\mu_2} = \frac{1}{m_N} + \frac{1}{2m_N + m_\pi}\ .    \label{eq:3.5}
\end{equation}
In a similar manner, the $\pi N$ scattering in channel $\alpha$ in
Fig.~\ref{Fig.9}, is represented by the amplitude $t_\alpha(Q_1^\prime,
Q_3^\prime;E_2)$ which is given by
\begin{equation}
t_\alpha(Q_1^\prime,Q_3^\prime;E_2) =
g_\alpha(Q_1^\prime) \tau_\alpha(E_2) g_\alpha(Q_3^\prime)\ , \label{eq:3.6}
\end{equation}
where the energy available to the $\pi N$ system, $E_2$, is given by
\begin{equation}
E_2 = E + m_N - \frac{q_3^2}{2\mu_2} - \frac{{p'_3}^2}{2\mu_1}
    = E + m_N - \frac{q_1^2}{2\mu_2} - \frac{{p'_1}^2}{2\mu_1}\ .
\label{eq:3.7}
\end{equation}
In this way we have made use of the general
structure of the $\pi N$ amplitude in terms of a one-particle reducible (the
$s$-channel pole amplitude) and the one-particle irreducible (the non-pole
amplitude) to determine the three-body force contribution from
Fig.~\ref{Fig.1}. Although we have used a separable potential for the non-pole
amplitude, there is no reason why we could not have made use of the non-pole
contribution from a chiral Lagrangian~\cite{PJ91}, or a separable
approximation to such a chiral $\pi N$ amplitude~\cite{PA89}, other than the
fact that this would have imposed an additional complexity to the evaluation of
such an amplitude.

Having defined the basic ingredients required to  calculate the contribution
from the diagram in Fig.~\ref{Fig.1}, we turn our attention  to the practical
problem of calculating the overall contribution of the diagram in
Fig.~\ref{Fig.1}. The contribution from the process whereby nucleon 1 emits a
pion that scatters off nucleon 2 in the $\pi N$ channel $\alpha$, and then
gets absorbed on nucleon 3 is given by the expression
\begin{eqnarray}
 W^\alpha_{\beta_1 \beta_3} (p_1,q_1,p_3,q_3;E) &=&
\langle f^R(E_1) | Q_1 \rangle
\langle p_1 q_1 | G_{\pi NNN}(E) | p_1' q_1 \rangle
\langle Q_1' | g_\alpha \rangle
\tau_\alpha(E_2) \, \Gamma_{\beta_1\beta_3}       \nonumber \\
&&\quad\times\  \langle g_\alpha | Q_3' \rangle
\langle p_3'q_3 | G_{\pi NNN}(E) | p_3 q_3 \rangle
\langle Q_3 | f^R(E_3) \rangle\ ,                             \label{eq:3.8}
\end{eqnarray}
where $\beta_1$ and $\beta_3$ represent the  set of quantum numbers that label
the three-body channels in the final and initial states, respectively. The
coefficient $\Gamma_{\beta_1 \beta_3}$ is a factor determined by the
transformation between the different Jacobi momenta in the three-nucleon
system. The four-body Green's function $G_{\pi NNN}(E)$ can be written in
terms of the $\pi NN$ Green's function, $G_{\pi NN}$, as
\begin{eqnarray}
G_{\pi NNN}(E) &=& G_{\pi NN}\left(E- \frac{q_i^2}{2\mu_2}\right) \nonumber \\
     &=& \left(E_i - m_N - \frac{Q_i^2}{2m_N} -
 \sqrt{Q_i^2 + m_\pi^2}\right)^{-1},\quad(i=1,3)  \ .        \label{eq:3.9}
\end{eqnarray}
This allows us to employ the methods developed for pion exchange  in the
$NN-\pi NN$ problem and to write the partial wave projection of the process
whereby nucleon 1 emits a pion that will scatter off nucleon 2 in channel
$\alpha$ as
\begin{eqnarray}
Z_{\alpha'_3,\beta'_3}^{j^\prime_3,t^\prime_3}
\left(p_3^\prime,p_3;E - \frac{q_3^2}{2\mu_2}\right) &=&
\langle g_\alpha|Q_3'\rangle\langle p_3'\,q_3 | G_{\pi NNN}(E) |p_3
\,q_3 \rangle \langle Q_3 | f^R(E_3) \rangle            \nonumber \\
          &=&\sum_{L,a,b}\,A^{L,a,b}_{\alpha'_3,\beta'_3}
\ p_3^{\ell_\pi-a+b}\ {p_3'}^{1+a-b} \ \rho_2^{a+b}          \nonumber \\
 &&\quad\times
\frac{1}{2}\int\limits_{-1}^1 dy  \, {{Q_3^\prime}^{-\ell_\pi}
\,g_\alpha(Q_3')\, f^R (Q_3;E_3)\,Q_3^{-1}
\over E_3 - m_N - {1\over 2m_N}Q_3^2  -\sqrt{Q_3^2+m_\pi^2} }
\, P_L(y)\ ,                                               \label{eq:3.10}
\end{eqnarray}
where $\ell_\pi$ is the relative $\pi N$ orbital angular momentum in the $\pi
N$ amplitude. Here $\beta'_3$ gives the set of quantum numbers for the
coupling scheme  $[(\pi N_1)_N,N_2]$ resulting in a total angular momentum
$j^\prime_3$ and total isospin $t^\prime_3$. In a similar manner
$\alpha'_3=(\alpha,\gamma_3)$ gives the set of quantum numbers for the
coupling scheme $[(\pi N_2)_\alpha, N_1]$ giving rise to the same total
angular momentum $j^\prime_3$ and total isospin $t^\prime_3$. The quantum
numbers of $N_1$ are given by $\gamma_3 = (\ell^\prime_3,S^\prime_3)$, where
$\ell^\prime_3$ is the orbital angular momentum of $N_1$ relative to the
center of mass of $(\pi N_2)$, and $S^\prime_3$ ( the corresponding channel
spin) is the sum of the total angular momentum ($j_\pi$) of $(\pi N_2)$, and
the spin ($s_1$) of nucleon $N_1$, i.e., ${\bf S}^\prime_3 = {\bf j}_\pi + {\bf
s}_1$. In writing the above expression for the one pion exchange amplitude, we
have made use of the fact that both ${\bf Q}_3$ and ${\bf Q}_3'$ can be written
in terms of the momenta ${\bf p}_3$ and ${\bf p}_3'$ as
\begin{equation}
{\bf Q}_3  = {\bf p}_3' + \rho_2\,{\bf p}_3 \ , \qquad
{\bf Q}_3' = {\bf p}_3 + \rho_2 \,{\bf p}_3' \ ,               \label{eq:3.11}
\end{equation}
where
\begin{equation}
\rho_2 = {m_N \over m_N + m_\pi}\ ,                            \label{eq:3.12}
\end{equation}
and $y=\hat{\bf p}_3\cdot\hat{\bf p}_3'$. The coefficients
$A^{L,a,b}_{\alpha'_3,\beta'_3}$ are those used in the partial  wave expansion
of the Faddeev equation for a separable potential~\cite{AG90}. For the case
when the $\pi N$ channel $\alpha$ corresponds to the $P_{33}$ partial wave,
Eq.~(\ref{eq:3.10}) gives the $(j'_3,t'_3)$ partial wave projection of the
$NN-N\Delta$ transition potential. In general, since we have restricted our
analysis to the case of separable non-pole $\pi N$ amplitudes, we can
interpret the $\pi N$ amplitude in each partial wave to be dominated by an
$N^*$ in which case $Z_{\alpha'_3,\beta'_3}^{j'_3,t'_3}$ can be considered as
the transition potential for $NN\rightarrow NN^*$. The difference
between the traditional $NN-N\Delta$ transition potential and the above result
in Eq.~(\ref{eq:3.10}) is the fact that the present transition potential is
energy dependent and the parameters of the potential are determined by the
$\pi N$ data rather than by the $NN$ data.

This interpretation of $Z_{\alpha'_3,\beta_3'}^{j'_3,t'_3}$ as a partial wave
projected
transition potential will allow us to regards the diagram in Fig.~\ref{Fig.1}
to
correspond to an initial state of three nucleons with nucleons 1 and 2 in the
channel
$\beta_3'$ going to two nucleons plus an $N^*$, followed by the $N^*$ coupling
to nucleon 3 to form a final state of nucleon 2 and 3 in channel $\beta_1'$
with nucleon 1 as spectator. This corresponds to the exchange of an $N^*$ and
will allow us to partial wave expand the three-body force in a similar manner
to the expansion of the one pion exchange diagram treated above. The resultant
partial wave expansion of the diagram in Fig.~\ref{Fig.1} is given as
\begin{eqnarray}
 W_{\beta_1, \beta_3}^\alpha(p_1,q_1,p_3,q_3;E) &=&
{1\over2}\sum_{\gamma_1,\gamma_3}\sum_{L,a,b}\ \int\limits_{-1}^1 dx\,
(-)^R\ {q_1^{\ell_3'-a+b}\,q_3^{\ell_1'+a-b}
\over {p_1'}^{\ell_1'} {p_3'}^{\ell_3'}}\  \rho_1^{a+b}\
A^{L,a,b}_{\beta_1,\beta_3}\,P_L(x)                   \nonumber \\
 & & \nonumber \\
&&\times\, Z_{\beta_1',\alpha'_1}^{j'_1,t'_1}
\left(p_1,p_1';E-\frac{q_1^2}{2\mu_2}\right)\,\tau_\alpha(E_2)\,
Z_{\alpha'_3,\beta_3'}^{j'_3,t'_3}
\left(p_3',p_3;E-\frac{q_3^2}{2\mu_2}\right)\ ,\label{eq:3.13}
\end{eqnarray}
where the phase, $R=j_\pi+t_\pi + \ell_3' +S_3' +t_3'$,  results from changing
the coupling scheme to maintain consistency with the definition of the Jacobi
coordinates. Here, $(j_\pi,t_\pi)$ are the total angular momentum and isospin
of the pion with nucleon 2, i.e., they define the $\pi N$ channel $\alpha$,
while $\ell_i'$ $(i=1,3)$ are the orbital angular momenta corresponding to the
momenta $p'_i$, e.g. $\ell'_3$ is the orbital angular momentum of $N_1$
relative to the center of mass of $(\pi N_2)$. In this case, we have written
the momenta ${\bf p}'_3$ and  ${\bf p}'_1$ in terms of the momenta ${\bf q}_3$
and
${\bf q}_1$ as
\begin{equation}
{\bf p}'_3 = {\bf q}_1 + \rho_1\,{\bf q}_3 \ ,\qquad
{\bf p}'_1 = {\bf q}_3 + \rho_1\,{\bf q}_1 \ ,                 \label{eq:3.14}
\end{equation}
where
\begin{equation}
\rho_1 = {m_N \over 2m_N + m_\pi } \ ,                         \label{eq:3.15}
\end{equation}
and $x=\hat{\bf q}_1\cdot\hat{\bf q}_3$. Here again the coefficient
$A^{L,a,b}_{\beta_1,\beta_3}$ is the coefficient resulting from the angular
momentum  recoupling, i.e.,  $[(N_1N_2)_{\beta'_3},N_3] \rightarrow N_1
+ N^*_2 + N_3 \rightarrow [(N_2 N_3)_{\beta'_1},N_1]$.

To compare the three-body force given in Eq.~(\ref{eq:3.13}) with that
resulting from the $NN-N\Delta$ coupled channel approach\cite{PRB91}, we note
that the expression in Eq.~(\ref{eq:3.13}) basically consists of a transition
potential $NN\rightarrow NN^*$ followed by the propagation of the $N^*$
quasiparticle and finally the transition potential $NN^*\rightarrow NN$.
Making use of the definition of $\tau_\alpha(E_2)$, ( Eqs.~(\ref{eq:2.3}) and
(\ref{eq:2.4}) ), we can write this quasi-particle propagator for the case when
$\alpha$ refers to the $P_{33}$ channel as
\begin{eqnarray}
\tau_\alpha(E_2) &=& \left[(E_2-E_\Delta)\langle g_\alpha|G_{\pi N}(E_2)G_{\pi
N}(E_\Delta)|g_\alpha\rangle\right]^{-1}\nonumber \\
 && \nonumber \\
 &=& {\left[\langle g_\alpha|G_{\pi N}(E_2)G_{\pi N}(E_\Delta)
|g_\alpha\rangle\right]^{-1}}\over
{\left[E + m_N -E_\Delta - \frac{q_3^2}{2\mu_2} - \frac{{p'}^2_3}{2\mu_1}
\right]}
                \ .                                          \label{eq:3.16}
\end{eqnarray}
In writing this equation we have made use of the fact that
$\tau^{-1}_\alpha(E_\Delta) = 0$,  where $E_\Delta= m_\Delta -
\frac{i}{2}\Gamma = 1230 - 50 i$~MeV is the position of the $\Delta(1230)$
resonance, and $\alpha$ refers to the $P_{33}$ partial wave. This illustrates
the fact that our quasi-particle propagator is the free $NN\Delta$ propagator
for which the  $\Delta$ has a width, and this width has the correct energy
dependence as dictated by unitarity and the experimental phase shifts in this
channel. Thus if we include the contribution to the three-body force from the
$P_{33}$ $\pi N$ channel only, we have effectively included, to lowest order,
the contribution from the three-body force resulting from the $NN-N\Delta$
coupling. However, our result differs from the standard definition of this
contribution to the three-body force\cite{PRB91} in that we have employed the
$\pi N$ data rather than the $NN$ data to fix the parameters of this force,
and our $\Delta$ is a proper resonance in the $\pi N$ system and not a real
particle as it is often considered. The other difference between this
approach and that used in the $NN-N\Delta$ coupled channel approach is in the
choice of the pion propagator in the ``transition potential''. In the
$N\Delta$ coupled channel approach, the pion propagator is taken to be
$(k_\pi^2 + m_\pi^2)^{-1}$, which corresponds to the Feynman propagator with
the nucleons on-mass-shell. This propagator has no energy dependence. On the
other hand we have chosen the standard non-relativistic four particle Green's
function with a relativistic expression for the pion kinetic energy. This is
consistent with four-particle unitarity and is equivalent to a time ordered
propagator, and to that extent we have only one time order for our pion
exchange. Finally, we should note that the $NN-N\Delta$ coupled channel
approach treats the contribution of the $\Delta$ to all orders and as a
result includes the dispersive contribution which to a large extent cancels
the contribution of the three-body force.

The recent three-nucleon force results, reported by Pe\~na~{\it et
al.}\cite{PSSK93},  treat the $\Delta$ as a $\pi N$ resonance. To that
extent, the $\pi N$ amplitude in the $P_{33}$ channel is similar to that
presented here, in that the mass and width of the $\Delta$ have energy
dependence as dictated by $\pi N$ scattering data. Pe\~na~{\it et al.} have the
additional advantage that they not only have included the $\Delta$
contribution to the three-body force to all orders, but have also included the
dispersive contribution. However, by restricting their Hilbert space to $N$
and $\Delta$ and turning off all interaction in the pionic part of the
Hilbert space, they have not included the contribution of the $\pi N$
amplitude in other than the $P_{33}$ partial wave.
They find that the contribution of the non-resonant $\pi N$ amplitude is
very small as a result of the fact that the pionic component of the
three-nucleon wave function is small.

We now try to relate our three-body force with that used in the TM approach.
In the TM approach, the final $\pi N$ amplitude is written in terms of the pion
momenta, where as we use the $\pi N$ relative momenta ${\bf Q}'_1$ and ${\bf
Q}'_3$~\cite{Foot4}. To achieve their result, we recall that the partial wave
expansion of the $\pi N$ amplitude is given by
\begin{equation}
\langle{\bf Q}|t(E)|{\bf Q}'\rangle =
\sum_{\alpha} P_\alpha(\hat{\bf Q},\hat{\bf Q}')
\ t_\alpha(Q,Q';E)           \ ,                   \label{eq:3.17}
\end{equation}
where $\alpha = (\ell,j,T)$ are the quantum numbers  corresponding to the
orbital and total angular momentum and isospin of the $\pi N$ system, i.e.
$\alpha$ labels the different partial waves. The partial wave projection
operator
$P_\alpha(\hat{\bf Q},\hat{\bf Q}')$ is defined as
\begin{equation}
P_\alpha(\hat{\bf Q},\hat{\bf Q}') = P_T \sum_m \,
\langle\hat{\bf Q}|{\cal Y}_{\ell jm}\rangle\,
\langle{\cal Y}_{\ell jm}|\hat{\bf Q}'\rangle\ ,            \label{eq:3.18}
\end{equation}
with $P_T$ the projection operator for a given isospin channel, and
$\langle\hat{\bf Q}|{\cal Y}_{\ell jm}\rangle$ the eigenstates of the
orbital and total angular momentum of the $\pi N$ system. If we now write the
angular momentum projection operator in terms of ${\bf Q}\cdot{\bf Q}'$ and
${\bf Q}\times{\bf Q}'\cdot {\bbox{\sigma}}$ and the isospin projection
operator in terms of the Pauli isospin operator, we can write our
$\pi N$ amplitude, assuming the pion scatters of nucleon 1, in the form
presented by the
Tucson-Melbourne formulation as
\begin{equation}
t^{TM}({\bf Q},{\bf Q}') = ({\bbox{\tau}}_2\cdot{\bbox{\tau}}_3)\,
\left[ a + b {\bf Q}\cdot{\bf Q}' + c(Q^2 + {Q'}^2)\right]
+ d({\bbox{\tau}}_1\cdot({\bbox{\tau}}_3\times{\bbox{\tau}}_2))
({\bbox{\sigma}}_1\cdot({\bf Q}\times{\bf Q}') )  \ .      \label{eq:3.19}
\end{equation}
For the factors $a$, $b$, $c$ and $d$ to be constant,  as required by the
Tucson-Melbourne definition of the three-body force, we have to make the
following approximations:
(i)~Since the TM potential is derived from the off-mass-shell $\pi N$ amplitude
$T(\nu,t,q,q')$, where $t$ is the Mandelstam variable and $q$ $(q')$ is the
four momentum of the initial (final) pion, by expanding the amplitude about
$\nu=0$,
we need to determine the corresponding approximation for our off-energy-shell
$\pi
N$ amplitude. For the off-mass-shell amplitude the nucleon pole ($s=m_N^2$)
traces a
curve in the $\nu-t$ plane that crosses the $\nu$-axes close to
$\nu=0$\cite{C86}.
Since we find it difficult to directly relate the variables that in the
off-mass-shell
amplitude with the corresponding variables in off-energy-shell amplitude, we
have
chosen the position of the nucleon pole, i.e. $E=m_N$ or $s=m_N^2$, to be the
closest
approximation to $\nu=0$. In this way we approximate the energy in our
amplitude
to
be the nucleon mass, {\it i.e.}, $\tau(E_2)\rightarrow\tau(m_N)$.
(ii)~The separable potential form factors
$g_\alpha(Q)$ have to be expanded in a power series in the momentum $Q$,
keeping those terms such that the final amplitude does not have any powers of
the momentum higher than the momentum squared. With these approximations, the
Thomas separable potential, in conjunction with the potential $PJ$ in the
$P_{11}$ channel, can be written in the above form with the constants $a$,
$b$, $c$ and $d$ given in Table~\ref{Table.3}. Included in the table are also
the corresponding parameters from the Tucson-Melbourne potential\cite{CP93}.

In next section, we will present our numerical result and show:
(i) How the  contribution to the three-body force from each $\pi N$ partial
wave
depends on the energy in the $\pi N$ amplitude.
(ii) How the three-nucleon force is sensitive to  the choice of the $\pi NN$
form
factor which is determined as the  residue of the $P_{11}$ pole term.
Though we have not included all the diagrams that would
contribute to the three-body force, we have included the most important
contribution, which will allow us to examine these effects.

\section{Numerical Results}\label{sec.4}

Having defined our three-nucleon force in terms of the $\pi N$ potential
whose parameters have been adjusted to fit the $\pi N$ data, and in
particular, the phase shifts up to the threshold for pion production, we turn
our attention in this section to the calculation of the contribution of this
force to the binding energy of the three-nucleon system. As a first
calculation with a three-body potential that includes the energy dependence of
the $\pi N$ amplitude, we have chosen to use the first order perturbation
theory
to calculate this three-body force contribution. Therefore, we can write our
three-body force as
\begin{eqnarray}
W &=& \sum_{{i,j=1}\atop{i\neq j}}^3\ W(j,i) \nonumber \\
  &=& \sum_{{i,j=1}\atop{i\neq j}}^3 \sum_\alpha \
W_{j,i}^\alpha({\bf p}_j,{\bf q}_j;{\bf p}^\prime_i,{\bf q}^\prime_i)
 \ ,                                                     \label{eq:4.1}
\end{eqnarray}
where $j$ in the sum refers to the nucleon that emits  the pion, and $i$
refers to the nucleon that absorbs the pion. Here, $\alpha$ refers to the $\pi
N$ partial wave used to calculate the three-body force. Since the
three-nucleon wave function used is a solution of the Faddeev equations, we
can write this wave function as the sum of three components, i.e.
\begin{eqnarray}
|\Psi\rangle &=& |\psi_1\rangle + |\psi_2\rangle + |\psi_3\rangle \nonumber \\
             &=& \left[ 1 + (123) + (132)\right]\,|\psi_1\rangle\ ,
                                                         \label{eq:4.2}
\end{eqnarray}
where we have written the total wave function in terms of the elements of the
permutation operators. Making use of the properties of  the permutation group,
we can write the total contribution to the binding energy from this three-body
force as
\begin{equation}
\Delta E^{(3)} =6\,\langle\Psi|\,W(1,3)\,|\Psi\rangle \ .       \label{eq:4.3}
\end{equation}
The three-body potential $W(1,3)$ can now be partial wave expanded in terms
of the partial wave potential given in Eq.~(\ref{eq:3.13}) as
\begin{equation}
\langle{\bf p}_1{\bf q}_1|\,W(1,3)\,|{\bf p}_3{\bf q}_3\rangle =
\sum_{\alpha}\,\sum_{\beta_1\beta_2}\ \langle{\bf \hat{p}}_1{\bf \hat q}_1|
\beta_1\rangle\,W^\alpha_{\beta_1\beta_3}(p_1,q_1,p_3,q_3;E)\,
\langle\beta_3|{\bf \hat p}_3{\bf \hat q}_3\rangle\ ,           \label{eq:4.4}
\end{equation}
where $\beta_i$ defines the three-body partial wave quantum  number in which
nucleon $i$ is the spectator. In Eq.~(\ref{eq:4.4}), $\alpha$ refers to the
$\pi N$ partial wave that contributes to the three-body potential. In a
similar manner we have to expand the three-nucleon wave function in terms of
the angular momentum and isospin bases
$\langle{\bf \hat{p}}_1{\bf \hat q}_1| \beta_1\rangle$. This is given by
\begin{eqnarray}
\langle{\bf p}_1{\bf q}_1|\Psi\rangle &=&
\sum_{\beta_1}\ \langle{\bf \hat p}_1{\bf \hat q}_1|\beta_1\rangle\,
\langle\beta_1;p_1 q_1|\Psi\rangle \nonumber \\
&=&\sum_{\beta_1=1}^{N_2}\ \langle{\bf \hat p}_1{\bf \hat q}_1|\beta_1\rangle\,
\left[\langle p_1 q_1|\psi_{\beta_1}\rangle +
\sum_{j=2,3}\ \sum_{\beta_j=1}^{N_1}\ \langle\beta_1|\beta_j\rangle\,
\langle p_j q_j|\psi_{\beta_j}\rangle\right]\ .               \label{eq:4.5}
\end{eqnarray}
In Eq.~(\ref{eq:4.5}) the sum over $N_1$ extends to the number of  three-body
channels included in the solution of the Faddeev equations which in turn is
determined by the number of two-body $NN$ channels included. On the other
hand, the sum over $N_2$ is an infinity sum which we have truncated for
practical calculations. For $N_2>N_1$ the first term $\langle
p_1q_1|\psi_{\beta_1}\rangle$ contributed only to the first $N_1$ terms in the
$N_2$ sum. This restriction is the result of truncating the number of $NN$
partial waves included in the solution of the Faddeev equations. We will test
the convergence of our final results to both the sum over $N_1$ and $N_2$.
Making use of the partial wave expansion for the three-body potential and the
three-nucleon wave function, we can write the total contribution of the
three-body force to the binding energy of the triton as
\begin{eqnarray}
\Delta E^{(3)} &=& 6\ \sum_\alpha\ \sum_{\beta_1\beta_3}\
\int\limits_0^\infty\,dp_1\,p_1^2\,\int\limits_0^\infty\,dp_3\,p_3^2\,
\int\limits_0^\infty\,dq_1\,q_1^2\,\int\limits_0^\infty\,dq_3\,q_3^2\nonumber\\
& & \qquad\times \langle\Psi|p_1 q_1;\beta_1\rangle\,
   W^\alpha_{\beta_1\beta_3}(p_1,q_1,p_3,q_3;E)\,
   \langle\beta_3;q_3 p_3|\Psi\rangle\ ,                      \label{eq:4.6}
\end{eqnarray}
where $E$ is the energy of the three-nucleon system as determined by the
solution of the Faddeev equations for a given two-nucleon interaction, and the
partial wave three-body potential, $W^\alpha_{\beta_1\beta_3}$, as given in
Eq.~(\ref{eq:3.13}).

In the present investigation we have chosen to use the Paris nucleon-nucleon
potential~\cite{LL80}.
The Paris potential is energy independent while our three-nucleon
force has been derived to be energy-dependent.
To that extent, our two- and three-body potentials are not consistent in
that they are not derived from the same Lagrangian.
However, since we are
examining the energy dependence of the three-body force for each
$\pi N$ partial wave, we hope that the present perturbative calculation may
allow us
to  gain some insight into this problem.
To simplify the construction of the
three-nucleon wave function needed to evaluate the integrals in
Eq.~(\ref{eq:4.6}), we have chosen the separable representation of the Paris
potential~\cite{HP84}. This representation has been tested for the
three-nucleon observables with considerable success~\cite{PK91}. In
Table~\ref{Table.4}, we give the rank of the separable expansion we have
chosen. This choice was dictated by the requirement that we should reproduce
the binding energy of the triton and the different percentages of $S-$, $S'-$,
and $D-$state probability for the triton. However, before we compare our
results with the coordinate space calculation for the Paris potential, we
present in Table~\ref{Table.5} the convergence of these quantities as we
increase the number of three-body channels, $N_1$, in the solution of the
Faddeev equations. From this table we can see that as far as the three-nucleon
wave function is concerned, the 18 channel Faddeev equations give good
convergence for all quantities. These 18 channels in the Faddeev equations
correspond to the truncation of the $NN$ interaction to include all two-body
partial
waves with total angular momentum less than or equal to two, including the
coupled
$^3P_2-^3F_2$ $NN$ channels. To justify the use of the separable expansion to
the Paris
potential, PEST~\cite{HP84}, we compare, in Table~\ref{Table.6}, our results
for
the 18
channel calculation with the corresponding results based on the coordinate
space solution of the Faddeev equations for the Paris potential. Taking the
difference between the two coordinate space calculations as a measure of the
numerical uncertainty in the solution of the Faddeev equations, we have good
agreement with previous results for the Paris potential. This suggests that
the three-nucleon wave function resulting from the PEST approximation is
comparable to that resulting from a solution of the Faddeev equations in
coordinate space for the exact Paris potential.

Having established the fact that the three-nucleon wave function generated by
the separable expansion to the Paris potential is of comparable quality to
that resulting from the solution of the Faddeev equations in coordinate space,
we turn our attention to the convergence of the contribution of the three-body
force to the number of three-body channels in the solution of the Faddeev
equations, $N_1$, and the number of three-body channels included in the
partial wave expansion of the wave function, $N_2$.
For this study we make use of the $P_{11}$ potential $PJ$ of McLeod and Afnan
\cite{MA85}. In Table~\ref{Table.7} we present
the contribution to the binding energy from the more important $\pi N$ partial
waves
for 5, 10, and 18 channel Faddeev calculations. In all cases we have taken 18
partial waves for the expansion of the wave function. All energies in
Table~\ref{Table.7} are in keV. We note at this stage that although the
contribution
to the binding energy is small, the 18 channel Faddeev calculation has
converged,
while the 5 channel calculation gives an incorrect result. With the 18
channel Faddeev calculation, we tested the convergence of our result to the
number of terms in the partial wave expansion of the three-body wave
function, $N_2$. From the results in Table~\ref{Table.8} we may conclude that
$N_2=18$  is sufficient to give us a 1~keV accuracy for the contribution from a
given
$\pi N$ partial wave. If the need arises we might have to resort to more terms
in
the partial wave expansion of the wave function.

In Table~\ref{Table.9} we present the contribution of the three-body force to
the binding energy of the triton from the different $\pi N$ partial waves for
two different $P_{11}$ potentials. Here we have taken $N_1=N_2=18$, with all
energies given in keV. The most surprising result of our calculations is the
overall small contribution of the three-body force. From the results in
Tables~\ref{Table.7} and \ref{Table.8}, it is clear that the inclusion of more
three-body partial waves in the wave function expansion and the solution of
the Faddeev equation will not change the results substantially. Before we
address the origin of this small contribution from our three-nucleon force it
is interesting to note: (i)~The comparable contribution from the $S-$ and
$P-$waves $\pi N$ partial wave, and in particular the large contribution of
the $S_{31}$ compared to the $S_{11}$. This suggests that we need to include
both $S-$ and $P-$wave $\pi N$ amplitudes into the calculation. Furthermore,
models based on the dominance of the $\Delta(1230)$ resonance might not be
valid since they neglect the contribution from the $S_{31}$ and $P_{11}$
partial waves. In fact, in the present formulation, the contribution of the
$P_{33}$ partial wave amplitude is for $\pi N$  energies below the nucleon
pole, some 300~MeV below the $\Delta$ resonance. (ii)~There is a cancellation
between the $S-$ and $P-$wave $\pi N$ contributions requiring a consistent
treatment of both sets of partial wave amplitudes. (iii)~The $P-$ wave
contribution comes equally from the non-pole part of the $P_{11}$, and the
$P_{33}$ partial wave amplitudes. This is despite the fact that the overall
$P_{11}$ phase shifts are small when compared with the $P_{33}$ phase shifts.
However, if we recall that it is the non-pole part of the amplitude that
contributes to the three-body force, and this non-pole part, on its own, has
phase shifts that are comparable to those in the $P_{33}$
channel~\cite{MF81,MA82}, then the results  reported in Table~\ref{Table.9}
are not surprising. Finally, if we compare the results for the two potentials,
we find that the potential $M1$ gives a larger contribution to the binding
energy than the potential $PJ$. To understand this difference, we compare the
dressed form factors for these two potentials in Fig.~\ref{Fig.10}. Here we
observe that the potential $M1$ has a harder form factor than the potential
$PJ$, i.e., the dressed form factor for potential $M1$ is greater than the
corresponding form factor for the potential $PJ$ for large $k$. This is
consistent with results of the fact that the three-body force contribution to
the binding energy increases as the form factor gets harder. We will come back
to this point later in our discussion when we consider the role of the $\pi
NN$ form factor in the contribution of the three-nucleon force to the binding
energy of the triton.

We now turn to the question of why the contribution of this three-body force
is small. From Table~\ref{Table.3}, we may expect the maximum difference
between

our prediction and the TM result for the three-body force contribution
to be at most an order of magnitude, but not three orders of magnitude.
Since the unique feature of the present calculation is the inclusion
of the energy dependence in both the $\pi NN$ form factor and the $\pi N$
amplitude, we will commence by turning this energy dependence off. We will
also concentrate on those $\pi N$ partial waves that give a substantial
contribution to the three-body force. We will restrict our results to
$N_1=N_2=18$. As a first approximation, denoted (i) in Table~\ref{Table.10},
we fix the energy in the $\pi NN$ form factor to be
the position of the nucleon pole, i.e.
\begin{equation}
f^R(k;E) \rightarrow f^R(k;m_N)  \ .                       \label{eq:4.7}
\end{equation}
Although this approximation changes our final result by increasing the
contribution of the three-body force to the binding energy, the magnitude of
the increase is not substantial because the cancellation between the
repulsive $S_{31}$ and the attractive $P_{11}$ and $P_{33}$ contributions is
still present. In particular, we note that both the attractive and repulsive
contributions have increased in magnitude. We next take the energy dependence
in the $\pi N$ amplitude to be the position of the nucleon pole
(approximation (ii)), i.e.
\begin{equation}
\tau_\alpha(E) \rightarrow \tau_\alpha(m_N)\ .                 \label{eq:4.8}
\end{equation}
In this case, we have increased the total contribution of the three-body
force to the binding energy by an order of magnitude as compared to the exact
result. This substantial increase in binding is mainly due to the fact that
the $S_{31}$ contribution is reduced in magnitude while the $P_{11}$ and
$P_{33}$ contributions have increased, thus reducing the cancellation between
the attraction and repulsion when compared with the exact calculation. To
understand why the contribution of the $S_{31}$ partial wave is suppressed as
the energy in $\tau_\alpha$ is increased (i.e. brought closer to the $\pi N$
threshold, see Fig.~\ref{Fig.3}), we recall from Eqs.~(\ref{eq:2.3}) and
(\ref{eq:2.4}) that $\lambda = +1$ for repulsive potentials such as the
$S_{31}$, while $\lambda= -1$ for attractive potentials such as the $P_{33}$.
On the other hand, $\langle g_\alpha|G_{\pi N}(E)|g_\alpha\rangle$ is negative
for $E<(m_\pi+m_N)$ and increases in value as we approach the $\pi N$
threshold from below. Thus for attractive potentials there is a cancellation
in the denominator of $\tau_\alpha(E)$ giving rise to an increase in the value
of $\tau_\alpha(E)$ as $E$ approaches the threshold. On the other hand, for
repulsive potentials, e.g. the $S_{31}$ channel, the value of $\tau_\alpha(E)$
decreases as we approach the elastic threshold from below, resulting in a
suppression of the repulsive contribution. To demonstrate the validity of
this argument, we have proceeded to change the energy in the $\pi N$
amplitude, to be the threshold for $\pi N$ scattering, i.e.,
\begin{equation}
\tau_\alpha(E) \rightarrow \tau_\alpha(m_N+m_\pi)   \ .     \label{eq:4.9}
\end{equation}
The result of this approximation is labeled (iii) in  Table~\ref{Table.10}.
Here, we observe that the contribution of the $S_{31}$ is further reduced,
while the $P_{11}$ and $P_{33}$ contributions are increased in magnitude,
giving even more attraction. We now fix the energy of both the $\pi NN$ form
factor and the $\pi N$ scattering amplitude to the nucleon pole, i.e. $E=m_N$,
(i) \& (ii) in Table~\ref{Table.10}. This gives even more binding than fixing
the energy in either the $\pi N$ scattering amplitude or the $\pi NN$ form
factor. Finally, we can increase binding further by fixing the energy of the
$\pi NN$ form factor at the nucleon pole, while the energy in the $\pi N$
amplitude is taken to be the threshold energy, (i) \& (iii) in
Table~\ref{Table.10}. From the above analysis we may conclude that it is the
energy dependence in both the $\pi NN$ form factor and the $\pi N$ amplitude
that has substantially reduced the contribution of the three-nucleon force to
the binding energy of the triton, and this reduction is a result of the
cancellation between the repulsive $S_{31}$ contribution and attractive
$P_{11}$ and $P_{33}$ contributions.
However, the approximation of fixing the energy in the $\pi N$ amplitude
and $\pi NN$ form factor does not give us a sufficiently large contribution
which is comparable with the result for the TM potential.

Finally, to fully understand the origin  of this small contribution to the
binding energy from the three-nucleon force, we turn our attention to the form
factors used in the separable potential and the form factor in the $\pi NN$
vertex. It is common practice to take the $\pi NN$ form factor in $NN$
scattering to be either of a dipole or monopole form. Therefore, as a first
step in examining the sensitivity of our final three-body force contribution
to the binding energy, we replace the $\pi NN$ form factor $f^R(k;E)$ by a
monopole, i.e.
\begin{equation}
f^R(k;E) \rightarrow \frac{f^R(k;m_N)}{k}\Bigg|_{k=0}\,k\,
 F_0(k)\ ,                                                 \label{eq:4.10}
\end{equation}
where the monopole form factor $F_0(k)$ is given by
\begin{equation}
F_0(k) = \frac{\Lambda^2}{\Lambda^2 + k^2}                 \label{eq:4.11}
\end{equation}
with the cutoff mass $\Lambda$, varied. In Table~\ref{Table.11} we compare
exact results for the $P_{11}$ potential $PJ$~\cite{MA85} with the result for
the approximations in Eqs.~(\ref{eq:4.7}) and (\ref{eq:4.8}),  referred to as
(i)~\&~(ii), and the approximations in Eqs.~(\ref{eq:4.8}) and (\ref{eq:4.10})
with $\Lambda = 400$ and $800$~MeV. The results in lines 2, 3 and 4 of
Table~\ref{Table.11} have the energy in the $\pi N$ amplitude fixed to be the
energy at the nucleon pole, i.e. $E_{\pi N} = m_N$. Here we observe that the
final total contribution of the three-body force increases by an order of
magnitude when the energy in both the $\pi NN$ form factor and $\pi N$
amplitude is fixed at the nucleon pole. There is a further order of magnitude
increase when the $\pi NN$ form factor is replaced by a monopole form factor
with a cutoff mass of 400 and then 800~MeV. In fact, half of this increase is
achieved when the cutoff mass is increased from 400 to 800~MeV. This
establishes the sensitivity of our result to the choice of $\pi NN$ form
factor. In this case it is interesting to observe that the $S_{31}$ and
$P_{11}$ contributions to the three-nucleon force almost completely cancel,
leaving the $P_{33}$ contribution to be approximately the total contribution.
Thus in this approximation, the contribution to the three-nucleon force is
predominantly due to the channel in which the $\Delta(1230)$ dominates the
scattering amplitude. This result should be compared with the exact results,
line 1 of Table~\ref{Table.11}, where the $S_{31}$ contribution cancels the
sum of the $P_{11}$ and $P_{33}$ contributions.

In the spirit of the TM approach, the $\pi N$ amplitude is expanded to
lowest order in $q\over m_N$  and a monopole  $\pi NN$ form factor is
introduced. This approximation may be implemented  by the following
replacement:
\begin{eqnarray}
f^R(k;E) &\rightarrow& \frac{f^R(k;m_N)}{k}\Bigg|_{k=0}\,k\, F_0(k)
\nonumber \\
\tau_\alpha(E)  &\rightarrow& \tau_\alpha(m_N)          \label{eq:4.12}
\\ g_\alpha(k)&\rightarrow& \frac{g_\alpha(k)}{k^\ell}\Bigg|_{k=0}
\ k^\ell\, F_0(k) \
,\nonumber
\end{eqnarray}
where $\ell$ is the angular momentum in channel $\alpha$ and the monopole
form factor $F_0(k)$ is defined in Eq.~(\ref{eq:4.11}).
However in our formulation, this replacement will destroy the fit to the
experimental $\pi N$ phase shifts. In
Table~\ref{Table.13}, we compare our exact result for the $P_{11}$ potential
$PJ$~\cite{MA85}, with the results of the approximation in Eq.~(\ref{eq:4.12})
with $\Lambda = 400$ and $800$~MeV.  The approximation in Eq.~(\ref{eq:4.12})
gives rise to an increase in the contribution of the $P$-waves substantially,
while the $S$-wave contribution remains relatively unchanged, and therefore
negligible.  In fact we can now adjust the cutoff mass $\Lambda$ to get the
difference between the experimental binding energy and the calculated
three-nucleon result for any of the two-nucleon interactions.

To understand this large change in the magnitude of the total contribution of
the three-body force when we introduce the form factor $F_0(k)$, we compare in
Fig.~\ref{Fig.10} the monopole form factor with $\Lambda =400$ and $800$~MeV
and
the
dressed $\pi NN$ form factor for the potentials $PJ$ and $M1$. Here we
observe that the form factor $F_0(k)$ with $\Lambda = 800$~MeV is almost a
factor of 3 larger than the dressed $\pi NN$ form factor at $k\approx
3$~fm$^{-1}$. Furthermore, this form factor comes raised to the power of two
in the case when only the $\pi NN$ form factor is replaced by a monopole form
factor, or a power of four when the approximations in Eq.~(\ref{eq:4.12}) are
implemented, i.e. we have a power of two from the $\pi NN$ vertices and another
power of two from the form factor of separable $\pi N$ amplitudes. Thus the
difference between the result of including $f^R(k,m_N)$ and
$F_0(k)$ should be roughly one order of magnitude in Tables~\ref{Table.11},
and two
orders of magnitude in Table~\ref{Table.13}. In the latter case we assume the
separable potential form factors have similar ranges to the dressed $\pi NN$
form
factor. This explains the drastic change in the contribution of the three-body
force
to the binding energy when we introduced the monopole form factor $F_0(k)$ into
our
calculation.

Finally, to get the closest approximation to the TM three-nucleon force, we
have modified the propagator for the exchanged pion by replacing our
propagator by the corresponding Feynman propagator, i.e.,
\begin{equation}
\frac{1}{Q_0 - \omega_Q}\rightarrow
\sqrt{2\omega_Q}\,\frac{1}{Q^2-m_\pi^2}\,\sqrt{2\omega_Q} \ ,\label{eq:4.13}
\end{equation}
where $\omega_Q = \sqrt{Q^2+m_\pi^2}$. To understand the difference between
these two propagators, we recall that the Feynman propagator has both a
positive energy and negative energy component, since
\begin{equation}
\frac{1}{Q^2 - m^2_\pi} = \frac{1}{2\omega_Q}\left[\frac{1}{Q_0 - \omega_Q} -
\frac{1}{Q_0 + \omega_Q}\right]\ .             \label{eq:4.14}
\end{equation}
and our choice of propagator, i.e. $(Q_0-\omega_Q)^{-1}$, corresponds to
taking the positive energy component of the Feynman propagator. However in any
energy independent approximation, the Feynman propagator reduces to $-
1/\omega^2_Q$. The substitution in Eq.~(\ref{eq:4.13}) gives us the last line
of Table~\ref{Table.13}, and results in the reduction of the contribution of
the three-body force so that a monopole form factor with a cutoff mass of
800~MeV will give a three-body force contribution of about 0.6~MeV, which is
consistent with the results reported in the literature for the TM
three-nucleon force.

Thus, to get a substantial contribution from the three-body force we have had
to make two approximations. (i) We have dropped the energy dependence of the
$\pi N$ amplitude and the $\pi NN$ form factor. (ii) We have modified the
off-shell behavior of the $\pi N$ amplitude by introducing the same monopole
form
factor in all partial waves, at the sacrifice of the fit to the experimental
data, in
order to get a substantial increase to the three-body force contribution.
Although
the first approximation is not justified, the second could be accepted on the
ground
that we have chosen the wrong off-shell behavior. In particular, we should
consider changes the cutoff mass in the bare $\pi NN$ form factor $f_0(k)$,
and make use of the Goldberger-Treiman~\cite{GT58} relation to constrain the
dressed $\pi NN$ form factor $f^R(k;E)$. This point is presently under further
investigation.

\section{Conclusion}\label{sec.5}

The main motivation for this investigation was to establish the relative
contribution of the different partial waves of the $\pi N$ amplitude in
determining the $\pi-\pi$ three-nucleon force and its contribution to the
binding energy of the triton. To achieve this, we made use of a separable
potential formulation of the $\pi N$ scattering, taking into consideration
that such a parameterization of the off-shell $\pi N$ amplitude has been
used in $\pi d$ scattering and in the derivation of the pion-nucleus optical
potential with considerable success. In particular, we maintained the
energy dependence of the $\pi N$ amplitude since that could effect the overall
contribution to the three-body force. More importantly, the cancellation
between the contribution of the different partial waves could be sensitive to
the inclusion of this energy dependence. Since the energy domain, important
to the determination of the three-nucleon force, is in the unphysical region
and below the position of the nucleon pole in the energy plane, we chose to
fit the scattering data closest to this region. In particular we chose to fit
the scattering lengths, the phase shifts up to the pion production threshold,
and the position and the residue of the $\pi N$ amplitude at the nucleon
pole. This parameterization allowed us to determine the $\pi NN$ form factor
and that part of the $\pi N$ amplitude that gives rise to the three-body
force.

{}From the numerical results of our calculation we can conclude that: (i)~The
energy dependence of both the $\pi NN$ form factor and $\pi N$ amplitude gives
a suppression of the contribution of Fig.~\ref{Fig.1} to the three-nucleon
force. Given the fact that this diagram has always been considered to give the
main contribution to this force, we can conclude that the three-nucleon force
for this $\pi N$ parameterization is small, and will not change substantially
if we include this three-body force in an exact calculation rather than in the
perturbative approach  used in the present investigation. (ii)~The inclusion
of the energy dependence gives rise to a substantial cancellation between the
contribution from the repulsive $S_{31}$ partial wave and the attractive
$P_{11}$ and $P_{33}$ partial waves. (iii)~The contribution from the $P_{33}$
partial wave is not as dominant as we would expect. In fact the attraction
comes equally from the $P_{11}$ and $P_{33}$, while the repulsion comes from
the $S_{31}$. This raises a question about the validity of including the
three-body force in terms of the $NN-N\Delta$ coupled channel approach while
neglecting the $S-$wave and the $P_{11}$  $\pi N$ contributions. (iv)~The
choice of the $\pi NN$ form factor, to be determined by the residue of the $\pi
N$ amplitude at the nucleon pole, is the other main reason for the reduction in
the overall magnitude of the three-nucleon force contribution to the binding
energy. The question of the possibility of choosing the cutoff mass in the
bare form factor $f_0(k)$ to be consistent with the $\pi N$ data, while
maintaining a substantial three-body force, will need further investigation.
Furthermore, the Goldberger-Treiman relation~\cite{GT58,CS90} should be used
to constrain the momentum dependence of the dressed $\pi NN$ form factor
$f^R(k;E)$.

To further substantiate the above conclusions, we may need to examine a number
of questions: (i)~How  sensitive are our results to the choice of the $\pi N$
interaction? In particular, would we get a small contribution from the
three-nucleon force, and specifically the diagram in Fig.~\ref{Fig.1}, if we
commenced with a chiral Lagrangian such as that used by Pearce and
Jennings~\cite{PJ91}? (ii)~Would our final results be substantially different
if this three-body force were to be included to all orders? Examination of the
perturbation series for the Paris potential with the TM three-body
force\cite{Bo86} suggests that higher order contributions are not negligible,
but the magnitude of the overall contribution does not change by more than a
factor of two. This further suggests that an exact calculation will not effect
our
final conclusion. (iii)~How important are the dispersive effects? The latest
results based on the $NN-N\Delta$ coupled channel\cite{PRB91} approach suggests
that there is a substantial cancellation between the three-body force
contribution and the dispersive effects in the $P_{33}$ channel. What happens
to this cancellation when other $\pi N$ partial waves are included? If this
cancellation is present for all $\pi N$ partial waves then it would lead to
further reduction in the three-body force.

\newpage

\section{ACKNOWLEDGMENT}

The authors would like to thank the Australian Research Council and Flinders
University Board  of Research for their financial support during the course of
this work. The authors would also like to thank J. Haidenbauer for supplying
them
with the parameters of the PEST potential, and D.R.~Lehman for the comparison
of
triton results for the PEST potential. Finally we would like to thank
B.F.~Gibson for supplying us with the results of the coordinate space
calculation for the triton with the Paris potential.

\newpage

\begin{figure}[h]
\caption{The contribution to the three-nucleon force.\label{Fig.1}}
\end{figure}

\begin{figure}[h]
\caption{The $NN$ complex energy plane with the unitarity cut and
the domain of integration when used in the Faddeev equation for the
three-nucleon system.\label{Fig.2}}
\end{figure}

\begin{figure}[h]
\caption{The $\pi N$ complex energy plane with the unitarity cut,
the $N$ and $\Delta$ pole position, and the region of integration
when calculating the three-body $\pi-\pi$ force.\label{Fig.3}}
\end{figure}

\begin{figure}[h]
\caption{The dressing of nucleons with the restriction on the Hilbert
space to include nucleons and up to one pion only. \label{Fig.4}}
\end{figure}

\begin{figure}[h]
\caption{The full dressing of nucleons on both sides of the $\pi NN$
vertex. \label{Fig.5}}
\end{figure}

\begin{figure}[h]
\caption{The lowest order terms resulting from the coupling of the $NNN$ to the
$\pi
NNN$ channels as given in Eq.(2.17).   (a) One pion exchange term,
${\cal V}_{OPE}$. (b) Dispersion term, ${\cal V}_{disp}$. And (c)
three body force, ${\cal V}_{3B}$.  \label{Fig.6}}
\end{figure}

\begin{figure}[h]
\caption{The nucleon pole contribution to the $\pi N$ potential.\label{Fig.7}}
\end{figure}

\begin{figure}[h]
\caption{The contribution of the crossed diagram and the contact
diagram to the $\pi N$ potential.\label{Fig.8}}
\end{figure}

\begin{figure}[h]
\caption{An illustration of the Jacobi momenta defined for the
three-body force.\label{Fig.9}}
\end{figure}

\begin{figure}[h]
\caption{Comparison of the dressed $\pi NN$ form factors for the potentials
$PJ$,
and $M1$ with a monopole form factor having cutoff mass of $\Lambda = 400,
800$~MeV.\label{Fig.10}}
\end{figure}

\newpage

\mediumtext
\begin{table}
\caption{Parameters of the McLeod-Afnan $P_{11}$ potentials.
}\label{Table.1}
\begin{tabular}{l|ccccccccc}
Pot.   &  $n_0$  & $n_1$ & $c_0$ & $c_1$ & $c_2$ &
$\alpha$ & $\beta_1$ & $\beta_2$ & $m_0$ \\
  &   &   &   &   &   &
 fm$^{-1}$ & fm$^{-1}$ & fm$^{-1}$ & fm$^{-1}$ \\ \tableline
$PJ$ & 2 & 1 &\dec 43.5646 &\dec 0.2907 &\dec 412.97 &\dec 3.8206 &\dec 1.2688
&\dec 5.181 &\dec 5.1574 \\
$M1$ & 1 & 2 &\dec 1.0726  &\dec 0.3433 &\dec   2.54 &\dec 2.7703 &\dec 1.4422
&\dec 2.1982 &\dec 5.4314 \\
\end{tabular}
\end{table}

\vskip 1 cm

\begin{table}
\caption{The scattering volume $a_{11}$, wave function renormalization $Z_2$,
and
the $\pi NN$ coupling constant at $k=0$ for the $P_{11}$ potentials.
The $f_{\pi NN}(0)$ is to be compared with a value of 0.2726 predicted by
the Goldberger-Treiman relation of 3\% change in $f_{\pi NN}(k)$ between
$k=k_0$ and $k=0$.}\label{Table.2}
\begin{tabular}{l|ccc}
Pot. & $a_{11}$ & $Z_2$ & $f_{\pi NN}(0)$ \\
     & ($m_\pi^{-3}$)  & & \\ \tableline
$PJ$   &\dec -0.0706  &\dec  0.8059 &\dec 0.2280 \\
$M1$   &\dec -0.0721  &\dec  0.7273 &\dec 0.2209 \\
\end{tabular}
\end{table}

\vskip 1 cm

\begin{table}
\caption{The parameters of the Tucson-Melbourne type $\pi N$  amplitude as
reported in Ref.\protect\cite{CP93}, and those extracted from the potential
of Thomas\protect\cite{Th76} and the $P_{11}$ potential
$PJ$\protect\cite{MA85}.}\label{Table.3}

\begin{tabular}{l|cccc}
Pot. & $a$ & $b$ & $c$ & $d$ \\ \tableline
Pe\~na \& Coon &\dec 1.03 &\dec -2.62 &\dec 0.91 &\dec -0.75 \\
Thomas \& $PJ$ &\dec 0.46 &\dec -4.66 &\dec 0.67 &\dec -2.47 \\
\end{tabular}
\end{table}

\vskip 1 cm

\begin{table}
\caption{The rank of the separable expansion to the Paris potential  for the
two-body $NN$ partial waves included in the calculation of the wave
function.}\label{Table.4}

\begin{tabular}{l|ccccc}
 &$^1S_0$ & $^3S_1$-$^3D_1$ & $^3P_0$,$^1P_1$,$^3P_1$ & $^3P_2$-$^3F_2$ &
$^1D_2$,$^3D_2$\\ \tableline
Rank & 3 & 4 & 2 & 3 & 2 \\
\end{tabular}
\end{table}

\vskip 1 cm

\begin{table}
\caption{The convergence of the binding energy of  the triton, the $S-$ $S'-$
and $D-$state probabilities for 5, 10 and 18 channel Faddeev calculations. In
calculating the wave function we have taken $N_2=146$. }\label{Table.5}

\begin{tabular}{c|cccc}
$N_1$ &  B.E. (MeV) & $P(S)$\% & $P(S')$\% & $P(D)$\% \\ \tableline
 5    &\dec 7.2659 &\dec 89.882 &\dec 1.652 &\dec 8.401 \\
 10   &\dec 7.0966 &\dec 90.274 &\dec 1.479 &\dec 8.182 \\
 18   &\dec 7.3175 &\dec 90.111 &\dec 1.430 &\dec 8.393 \\
\end{tabular}

\end{table}

\vskip 1 cm

\begin{table}
\caption{Comparison of our 18 channel results for the separable expansion
(PEST), with the results of coordinate space calculations of Los Alamos
(LA)~\protect\cite{FG88} and Sendai (S)~\protect\cite{SI86}.}\label{Table.6}

\begin{tabular}{l|cccc}
Model &  B.E. (MeV) & $P(S)$\% & $P(S')$\% & $P(D)$\% \\ \tableline
 PEST   &\dec 7.318 &\dec 90.111 &\dec 1.430 &\dec 8.393 \\
 LA     &\dec 7.388 &\dec 90.130 &\dec 1.395 &\dec 8.409 \\
 S      &\dec 7.56  &\dec 90.17  &\dec 1.32  &\dec 8.45  \\
\end{tabular}

\end{table}

\vskip 1 cm

\begin{table}
\caption{Convergence of the contribution of the three-nucleon force to
the binding energy of the triton from different $\pi N$ partial waves
and for different numbers of the three-body channels in the Faddeev
equations.  Here we have taken $N_2=18$. All energies are in
keV.}\label{Table.7}

\begin{tabular}{l|cccc}
 $N_1$ & $S_{11}$ & $S_{31}$ & $P_{11}$ & $P_{33}$  \\ \tableline
  5   &\dec - 5.5 &\dec 31.8 &\dec - 7.7 &\dec - 15.2 \\
 10   &\dec - 4.8 &\dec 26.5 &\dec - 7.9 &\dec - 15.0 \\
 18   &\dec - 4.8 &\dec 26.4 &\dec - 8.8 &\dec - 16.0 \\
\end{tabular}

\end{table}

\vskip 1 cm

\begin{table}
\caption{Convergence of the contribution of the three-nucleon force
to the binding energy from different $\pi N$ partial waves
as a function of $N_2$, the number of three-body channels in the partial
wave expansion of the wave function. The Faddeev equations are solved
with 18 channels.  All energies are in keV.}\label{Table.8}

\begin{tabular}{l|cccc}
$N_2$ & $S_{11}$ & $S_{31}$ & $P_{11}$ & $P_{33}$  \\ \tableline
 10  &\dec - 5.2 &\dec 37.1 &\dec - 7.2 &\dec - 8.5   \\
 18  &\dec - 4.8 &\dec 26.4 &\dec - 8.8 &\dec - 16.0  \\
 26  &\dec - 5.3 &\dec 26.2 &\dec - 9.5 &\dec - 16.1  \\
 34  &\dec - 5.3 &\dec 25.7 &\dec - 9.5 &\dec - 16.9  \\
\end{tabular}

\end{table}

\vskip 1 cm

\begin{table}
\caption{The contribution to the three-body force from different $\pi N$
partial
wave
amplitudes. The results are for two different choices of the $P_{11}$
potential.  All energies are in keV.}\label{Table.9}

\begin{tabular}{l|ccccccc}
Pot. & $S_{11}$ & $S_{31}$ & $P_{11}$ & $P_{31}$ & $P_{13}$ & $P_{33}$ &
Total\\
\tableline
 $PJ$  &\dec -4.8 &\dec 26.4 &\dec -8.8  &\dec -3.6 &\dec 4.5 &\dec -16.0 &\dec
-2.3
\\
 $M1$  &\dec -5.0 &\dec 28.9 &\dec -15.3 &\dec -2.1 &\dec 6.2 &\dec -22.1 &\dec
-9.4
\\ \end{tabular}

\end{table}

\begin{table}
\caption{ The effect of removing the energy dependence in the $\pi NN$ form
factor (i),
and the $\pi N$ amplitudes, (ii) and (iii). The results in this table are for
the
$P_{11}$ potential $PJ$, and all energies are in keV.}\label{Table.10}

\begin{tabular}{c|ccccccc}
Approx.     & $S_{11}$ & $S_{31}$ & $P_{11}$  & $P_{31}$ & $P_{13}$ & $P_{33}$
&
 Total     \\ \tableline
 exact      &\dec -4.8 &\dec 26.4 &\dec -8.8  &\dec -3.6 &\dec 4.5  &\dec -16.0
&
\dec -2.3  \\
 (i)        &\dec -5.9 &\dec 31.7 &\dec-13.0  &\dec -3.0 &\dec 5.7  &\dec -22.3
&
\dec -6.8  \\
 (ii)       &\dec -5.0 &\dec 23.5 &\dec-13.7  &\dec -3.6 &\dec 3.9  &\dec -28.2
&
\dec -23.1 \\
 (iii)      &\dec -5.7 &\dec 19.5 &\dec-19.6  &\dec -3.5 &\dec 3.6  &\dec -49.8
&
\dec -55.5 \\
(i) \& (ii) &\dec -6.1 &\dec 28.1 &\dec-20.1  &\dec -3.1 &\dec 5.0  &\dec -39.8
&
\dec -36.0 \\
(i) \& (iii)&\dec -7.0 &\dec 23.4 &\dec-28.7  &\dec -3.0 &\dec 4.7  &\dec -70.3
&
\dec -80.9 \\
\end{tabular}
\end{table}

\begin{table}
\caption{The effect of changing the $\pi NN$ form factor on the contribution
of the three-nucleon force to the binding energy of the triton. The energy in
the $\pi N$ amplitude is fixed at $m_N$. The comparison is between monopole
form factor with a cutoff mass $\Lambda = 400$ or $800$~MeV and $f^R(k,m_N)$.
Also included are the exact results which have the full energy dependence of
both the $\pi NN$ form factor and $\pi N$ amplitude. The total includes the
contributions from all $S$- and $P$-wave $\pi N$ amplitudes. All energies are
in keV.}\label{Table.11}

\begin{tabular}{c|ccccc}
$\Lambda$ & $S_{11}$ & $S_{31}$ & $P_{11}$ & $P_{33}$ & Total\\ \tableline
exact       &\dec  -4.8 &\dec  26.4 &\dec    -8.8 &\dec   -16.0 &\dec    -2.3
\\
(i) \& (ii) &\dec  -6.1 &\dec  28.1 &\dec   -20.1 &\dec   -39.8 &\dec   -36.0
\\
400         &\dec -12.5 &\dec  57.6 &\dec   -40.7 &\dec   -81.2 &\dec   -68.7
\\
800         &\dec -29.3 &\dec 147.5 &\dec  -204.1 &\dec  -364.0 &\dec  -395.1
\\
\end{tabular}
\end{table}

\begin{table}
\caption{The effect of changing the form factors on the contribution of the
three-nucleon force to the binding energy of the triton. Here, both the $\pi
NN$
form
factors$f^R(k;E)$ and the separable potential form factors $g_\alpha(k)$ are
replaced
by a monopole with a cutoff mass $\Lambda=400$ or $800$~MeV. The last line in
this
table corresponds to taking the `Feynman' propagator for the pion. The total
includes
the contributions from all $S$- and $P$-wave $\pi N$ amplitudes. All energies
are in
keV.}\label{Table.13}

\begin{tabular}{c|ccccc}
$\Lambda$ & $S_{11}$ & $S_{31}$ & $P_{11}$ & $P_{33}$ & Total\\ \tableline
exact     &\dec  -4.8 &\dec  26.4 &\dec    -8.8 &\dec   -16.0 &\dec    -2.3 \\
400       &\dec -15.0 &\dec  18.8 &\dec   -58.8 &\dec  -211.5 &\dec  -231.1 \\
800       &\dec -82.7 &\dec 113.2 &\dec -1146.7 &\dec -3172.5 &\dec -3897.2 \\
800$^*$   &\dec -36.0 &\dec  47.4 &\dec  -193.4 &\dec  -542.5 &\dec  -611.6 \\
\end{tabular}
\end{table}


\begin{references}
\bibitem{Foot1} Since the three nucleon wave function is determined within
the framework of the non-relativistic Faddeev equations, where all particles
are on-mass-shell and all amplitudes are off-energy-shell, we will consider
$\pi NN$ dynamics and therefore the $\pi N$ amplitude is in an
off-energy-shell formulation.
\end{references}
\end{document}